\documentclass[12pt,preprint]{aastex}
\usepackage{epsfig, array, amsmath, delarray}

\newcommand{\muas}{\mu as}

\shorttitle{Relativistic stellar aberration..}
\shortauthors{Turyshev}

\begin{document}

\title{Relativistic stellar aberration 
 for the \\ Space Interferometry Mission (2)}

\author{Slava G. Turyshev\altaffilmark{1}}
\affil{Jet Propulsion Laboratory, California Institute
of  Technology, Pasadena, CA 91109}

\begin{abstract} 
We address the issue of relativistic stellar aberration
requirements for the Space Interferometry Mission (SIM). 
Motivated by the importance of this issue for SIM, we have considered a
problem of relative astrometric observations of two stars separated  by 
angle $\theta$ on the sky with a  single baseline interferometer. 
While a definitive answer on the stellar aberration issue may be obtained only in
numerical simulations based on the accurate astrometric  model of the
instrument, one could still derive  realistic conclusions by  accounting for the
main expected properties of SIM. In particular, we have analysied how the
expected astrometric accuracy of  determination of positions, parallaxes and
proper motions will constrain the accuracy of the spaceraft navigation. We
estimated the astrometric errors introduced by imperfect metrology (variations
of the calibration term across  the tile of interest), errors in the  baseline
length estimations, and those due to orbital motion of the spacecraft. We also
estimate  requirements  on the data sampling rate necessary to apply on-board 
in order to correct for the stellar aberration. We have shown that the  worst
case observation scenario is realized for the motion of the spacecraft in the
direction perpendicular to the  tile. This case  of motion will provide  the
most stringent requirement  on  the accuracy of knowledge of the
velocity's magnitude.  We discuss the implication of the results obtained for
the future mission analysis.

\end{abstract} 

\keywords{astrometry; techniques: interferometric, SIM;  methods:
analytical; solar system; relativity}
 

\section*{Introduction}

One of the best promising methods of searching for a planetary systems  
around nearby stars is the high accuracy astrometric observations. SIM is
being designed to produce wealth of the astrometric data necessary to 
address exactly this problem. It is clear that the corresponding
astrometric signal due to a reflex motion of a target star will be at the
level of a few $\mu$as and smaller. The most successful searching
techniques will have to incorporate an intelligent way of data processing
almost at the astrometric noise level. Another important part of this
puzzle is development of the astrometric model for the instrument.
This model should include all the parameters necessary to account for 
different phenomena  affecting the light propagation, and are
due to  the interstellar media,  the solar system dynamics, as well as 
due to the motion of the free-flying interferometer itself. As a result,
the accuracy of astrometric observations expected with SIM, will require a
number of dynamical parameters to be precisely known. To do this, one will
have to be able to remove (or to correctly account for) the signatures of
all the known  effects in order to study the unknown phenomena.  This is
necessary for minimizing probability of a false detections caused, for
example,  by  aliasing  the  dynamics of objects in the solar system.
Actually, there  will be a number of  parameters introduced  by the
dynamics in  the solar system  with periods of the order of a few years.
One of such a  parameters is the three-dimensional vector of barycentric
velocity of the spacecraft.  Thus, a future astrometric model will have to
account not only for the effects due to the motion of the solar system
bodies, but also for those that are generated by the motion of the spacecraft 
and corresponding errors in the spacecraft's navigation.

In this paper we will discuss some important elements of dynamical model 
for the future astrometric observations with  SIM. Specifically, we will
address the issue of the relativistic stellar aberration.  The stellar
aberration is a very important problem for SIM. Effect of the barycentric
velocity of the spacecraft  is the  largest term in the SIM astrometric
model and will amounts to $\sim 20.5$ arcsec for the Earth-trailing orbit.
This effect is important not only for a wide angle astrometric observations,
but it will produce a measurable effect  for a pairs of widely  separated
stars even inside the tile (with diameter $\sim 15^\circ$). 
Additionally, a possible correlation of the astrometric observables with
the errors in the spacecraft's velocity sky-angles suggests that in order to
achieve the mission accuracy for a global astrometric observations of
$\sigma=4~\mu$as, one will have to control the barycentric velocity of the
spacecraft  throughout the entire mission and  account for  the
relativistic stellar aberration inside every single tile.

The outline of this paper is  as follows.  In Section \ref{sec:model} we
will present a simplified model for   differential astrometric observations 
with a free-flying interferometer with a single baseline. We will discuss
a model for absolute and differential astrometric measurements  and will
give a three-dimensional parameterization of main observables. We will
present the first order differentials, necessary to analyze the errors
propagation.  In Section \ref{sec:tile} we will discuss orientation of the
tile in the SIM nominal observing direction. We will present
the first order equations that are governing the propagation of astrometric
errors in the instrument. Specifically, we address the velocity knowledge
requirement, the data sampling rate and orbital position knowledge, and the
spacecraft acceleration accuracy constraint. In Section
\ref{sec:req} we will derive the set of the relativistic stellar aberration
requirements for SIM. We will conclude by  presenting our recommendations
for future attitude control for the SIM spacecraft.   In order to make 
access of the basic results  of this paper easier, we will present some
important (but lengthy!) calculations  in the Appendices.

\section{A model for the relativistic stellar  aberration}
\label{sec:model}

In this Section we will  derive a model necessary to analyze the
astrometric error budget for  differential astrometric observations
with a single-baseline interferometer.   To do this, we will present
estimates of the optical path difference (OPD) for a single-baseline
interferometer  in solar orbit.  Our derivations will be different
from the ones obtained earlier by the fact that for each tile we account
for effects of general three-dimensional motion of the interferometer,
for the errors in the components of the baseline vector $\vec b$, and for
the error in estimating the instrumental offset (or, calibration) term,
$c_0$.  Our model is very simple, but it could be easily expanded to 
accommodate a number of other important features of the instrument.

We address the problem of relativistic stellar aberration   from a general
standpoint. A ``toy'' astrometric model developed here will include only
the largest  relativistic contributions due to the orbital motion of the
interferometer around the sun.  A complete, fully-relativistic model for
the SIM observations, will have to account for a number of  physical
phenomena and should include a set of additional terms (and corresponding
parameters)  necessary for  different   astrometric   applications. More
specifically, the future model should include the terms
$\propto(\frac{v}{c})^3$ to account for higher orders of relativistic
stellar aberration; terms
$\propto~G{\cal M}/c^2$ (where $G$ is the gravity constant and ${\cal M}$ is
the mass of the  deflecting body) to account for the gravitational deflection
by the bodies of the solar system (primarily the Sun and the Jupiter). (For
general discussion of the general relativistic effects in the SIM
asdtrometric observations, see \cite{Turyshev02}) One also will need to
account for rigid-body rotational motion of the spacecraft, etc. All these
effects, together with a number of others, are out of the scope of the
present paper, but will be discussed elsewhere.  While the derivation of the
general expressions will be given in the reminder of the document, here we
will present only the main results obtained.

\subsection{\large Geometry of the problem}

Let us begin by deriving the  expression for   OPD
$\ell$ for an interferometer with a single baseline
$\vec{b}$ which is moving  with the velocity $v\ll c$ (for SIM this
ratio will of of the order $v/c\sim 10^{-4}$) with respect to some
reference frame (RF), for example the solar-system barycentric (SSB)
reference frame. In the interferometer's proper quasi-inertial reference
frame  OPD,
$\ell'$,  may be given as follows: 
\begin{equation}
\ell' =(\vec{b'}\cdot\vec{s'})+c'_0, 
\label{eqdel010}
\end{equation}
\noindent where $\vec{b'}$ is the baseline vector, $\vec{s'}$ is the 
direction to the observed source, and the last term  in this expression is
the instrument offset (or ``constant'' term), $c_0$, that can be calibrated
out. The same quantity may be presented in the coordinates of the SSB 
reference frame.   The corresponding transformation between the solar
system barycentric   frame and  the one of the  moving interferometer (with
velocity $\vec{v}$ with respect to  SSB) will necessarily  account for the
effect of  the relativistic stellar aberration  (for more details see
\cite{Turyshev98}).  To the first order in $\frac{v}{c}$, the aberrated
direction to the  source and the baseline vector may be expressed as
follows:  
\begin{eqnarray}
\vec{s'}&=& \vec{s}+\frac{\vec{v}}{c}
-\frac{\vec{s}(\vec{s}\cdot\vec{v})}{c}+ {\cal
O}\Big(G,(\frac{v}{c})^2\Big),\\ 
\vec{b'}&=&  \vec{b}+ {\cal O}\Big(G,(\frac{v}{c})^2\Big), 
\label{eqdel011}
\end{eqnarray}

\noindent where $\vec{s}$ and $\vec{b}$ are the direction to the observed
source  and the baseline vector as measured in SSB. The total
effect of this transformation  has the form
\begin{equation}
\ell =(\vec{b}\cdot\vec{s})\Big(1-\frac{(\vec{s}\cdot\vec{v})}{c}\Big)+
\frac{(\vec{b}\cdot\vec{v})}{c}+c_0+ {\cal O}\Big(G,(\frac{v}{c})^2\Big), 
\label{eqdel10}
\end{equation}
\noindent where we have assumed  that the instrument offset does not depend 
on the velocity of the spacecraft motion to the second order in
$\frac{v}{c}$, or, in more general terms, 
$c'_0=c_0+{\cal O}\Big(G,(\frac{v}{c})^2\Big)$. This issue may be correctly
addressed later when the accurate model of the instrument will be developed. 


\begin{figure}[ht]
 \begin{center}
\hskip -110pt 
    \psfig{figure=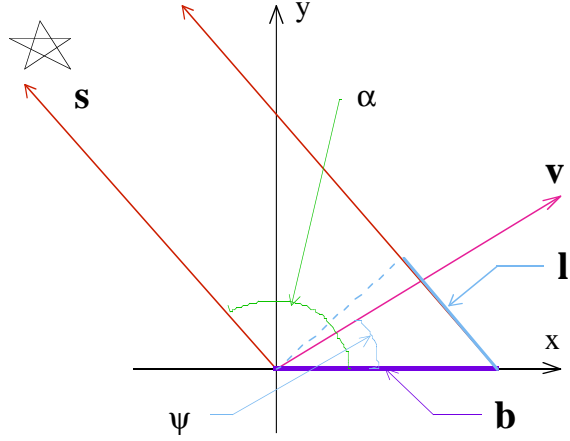,width=100mm,height=60mm}
     \caption{Two-dimensional geometry of the problem: absolute astrometry. 
      \label{fig:absolute_astrometry}}
 \end{center}
\end{figure}

 The simplified two-dimensional geometry of the problem  presented
in Figure \ref{fig:absolute_astrometry}.  This allows to express the
scalar products in   Eq.(\ref{eqdel10}) and re-write  this equation
in a more familiar form
{}
\begin{equation}
\ell  = b\cos\alpha\Big(1-\frac{v}{c}\cos(\alpha-\psi)\Big)+
\frac{bv}{c}\cos\psi +c_0. 
\label{eqdel12} 
\end{equation}

\noindent Such an expression may be used to model  absolute
astrometric observations with a moving interferometer, when the observed
sources are not in the same field of view, but rather widely separated from
each other. However, the wide-angle astrometric campaign with SIM will be
based on a set of differential (or relative) astrometric observations made
inside a set of 637 tiles covering the whole sky (\cite{Boden97,Swartz98}).
As a result, it  will   measure to a certain accuracy the angular distances
between a pairs of stars in the same field of view.  This necessitates the
derivation of an expression for the optical path difference, similar to
Eq.(\ref{eqdel12}),   which would take this fact into account. To the first
order in
$\frac{v}{c}$ this is a fairly easy task. Assuming that absolute
position for each  star is given by expression, similar to that of
Eq.(\ref{eqdel10}), namely:  
\begin{eqnarray}
\ell_1
&=&(\vec{b}\cdot\vec{s}_1)\Big(1-\frac{(\vec{s}_1\cdot\vec{v})}{c}\Big)+
\frac{(\vec{b}\cdot\vec{v})}{c}+c_{01},
\nonumber \\
\ell_2
&=&(\vec{b}\cdot\vec{s}_2)\Big(1-\frac{(\vec{s}_2\cdot\vec{v})}{c}\Big)+
\frac{(\vec{b}\cdot\vec{v})}{c}+c_{02}, 
\label{eqdel11a}
\end{eqnarray}  

\noindent one could write the corresponding  differential (or relative) 
OPD for a differential astrometric observations. To describe differential
measurements one will have to subtract one OPD from the other, namely
$\delta\ell=\ell_1-\ell_2$. The corresponding differential
OPD $\delta\ell$, will be given as follows:
{}
\begin{equation}
\delta\ell = 
(\vec{b}\cdot\vec{s}_1)\Big(1-\frac{(\vec{s}_1\cdot\vec{v})}{c}\Big)-
(\vec{b}\cdot\vec{s}_2)\Big(1-\frac{(\vec{s}_2\cdot\vec{v})}{c}\Big)+
\delta c_0 +{\cal O}\Big(G,(\frac{v}{c})^2\Big),
\label{eq:delmain} 
\end{equation}
\noindent where $\delta c_0$ is the differential ``constant'' term  
$\delta c_0=c_{01}-c_{02}$.

\subsection{Three-dimensional parameterization.}

In this subsection we will present the three-dimensional parameterization 
of all the vectors involved in the problem. This from of representation is
more useful  for practical applications, for example, to study
the orbital motion of the spacecraft; for analyzing the various
requirements on the spacecraft's velocity, rigid-body rotations, 
vibrations of the whole structure, etc.  

\begin{figure}[t]
 \begin{center}
    \psfig{figure=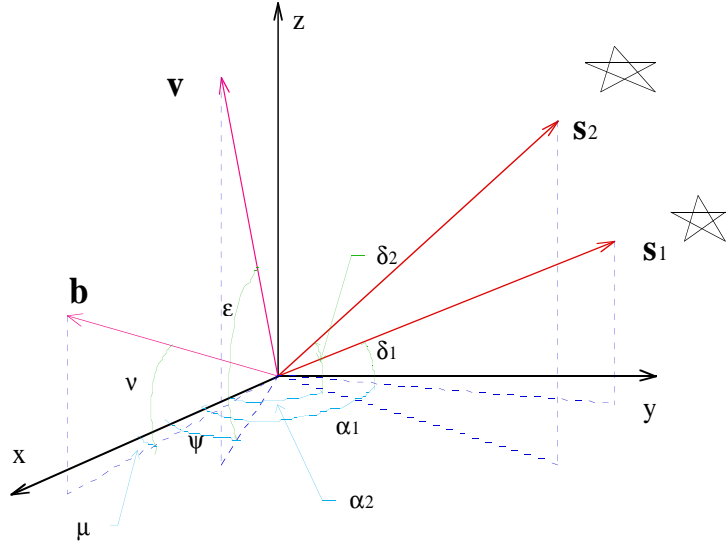,width=95mm,height=72mm}
     \caption{Geometry of the problem: differential astrometry. 
      \label{fig:differential_astrometry}}
 \end{center}
\end{figure}

The overall three-dimensional geometry of the problem   presented in 
Figure \ref{fig:differential_astrometry}. The different vectors 
involved are given in the  spherical coordinate
system  by their  magnitudes and  two corresponding   sky-angles:
\begin{eqnarray}
\vec{s}_1&=&\big(\cos\alpha_1\cos\delta_1,~\sin\alpha_1\cos\delta_1,
~\sin\delta_1\big),\nonumber\\[1mm]
\vec{s}_2&=&\big(\cos\alpha_2\cos\delta_2,~\sin\alpha_2\cos\delta_2,
~\sin\delta_2\big),\nonumber\\[1mm]
\vec{v}&=&v\big(\cos\psi\cos\epsilon,~\sin\psi\cos\epsilon,
~\sin\epsilon\,\big),\nonumber\\[1mm]
\vec{b}&=&b\big(\cos\mu\cos\nu,~\sin\mu\cos\nu,~\sin\nu\,\big). 
\label{eq:vect}
\end{eqnarray}

This parameterization allows one to present expression 
Eq.(\ref{eq:delmain}) in the following form:
 \begin{eqnarray}
\delta\ell &=& b\Big[\cos\nu\Big(\cos\delta_1\cos(\alpha_1-\mu)-
\cos\delta_2\cos(\alpha_2-\mu)\Big)+
\sin\nu\Big(\sin\delta_1-\sin\delta_2\Big)\Big]+\delta c_0-\nonumber\\[1mm]
&-&\frac{bv}{c}\Big[
\Big(\cos\nu\cos\delta_1\cos(\alpha_1-\mu)+\sin\nu\sin\delta_1\Big)
\Big(\cos\epsilon\cos\delta_1\cos(\alpha_1-\psi)+
\sin\epsilon\sin\delta_1\Big)-\nonumber\\[1mm]
&-&\Big(\cos\nu\cos\delta_2\cos(\alpha_2-\mu)+\sin\nu\sin\delta_2\Big)
\Big(\cos\epsilon\cos\delta_2\cos(\alpha_2-\psi)+
\sin\epsilon\sin\delta_2\Big)\Big].
\label{eq:del12a} 
\end{eqnarray}

This expression is quite difficult for analytical description, 
however, for the purposes of the present study,  it may be  simplified
(a more detailed analysis of this problem will be given in
the Appendix \ref{appa}). Thus, one may see that the angles of the baseline
orientation ($\mu,\nu$)  are significantly influencing the narrow-angle
astrometric observations. Angle $\mu$ corresponds to a shift of the
origin of  {\tt RA} for the \underline{\rm\emph{tile}} under consideration
and, without lost of generality, it may be omitted (e.q. set to be zero).  
The analysis of contribution of the {\tt DEC} angle $\nu$ is a little 
bit more complicated task. For the purposes of this paper we will assume
that this angle will be $\nu=0$ at the beginning of the 
\underline{\rm\emph{tile}} observation and will insignificantly 
deviate from this value during the experiment. This is true, because one
of the functions for the two guide interferometers in SIM 
will be  to provide a stable reference for the science
interferometer.  We, therefore will set both angles as
$\mu=\nu=0$. These parameters will have to be incorporated in the future
astrometric model for SIM, probably in a form of the rigid body
rotation/precession/nutation model of the spacecraft. At this moment, this
choice  is equivalent to choosing the  direction of the baseline vector
$\vec{b}$  coinciding with $x$-axis:
$\vec{b}=b(1,0,0)$ (see Figure \ref{fig:differential_astrometry}). As a
result, all vectors now will be counted 
\underline{from the baseline} and the expression (\ref{eq:del12a})  
may  now be presented in a  simpler  form, namely:   {} 

\begin{eqnarray}
\delta\ell &=&b\Big(\cos\delta_1\cos\alpha_1 -
\cos\delta_2\cos\alpha_2\Big)+ \delta c_0-\nonumber\\
&-&\frac{bv}{c} 
\Big[\cos\delta_1\cos\alpha_1 
\Big(\cos\epsilon\cos\delta_1\cos(\alpha_1-\psi)+
\sin\epsilon\sin\delta_1\Big)-\nonumber\\
&&~~-\cos\delta_2\cos\alpha_2 
\Big(\cos\epsilon\cos\delta_2\cos(\alpha_2-\psi)+
\sin\epsilon\sin\delta_2\Big)\Big].
\label{eq:del12b} 
\end{eqnarray}

The obtained expression may now be used to analyze the error propagation 
for the moving interferometer. To do this, one will have to expand this
expression in a Taylor series and keep only the first term of the
expansion. While the details of this calculations are given in
Appendix \ref{appa}, here we present only the final expression 
which is necessary to analyze  contributions of the chosen set of 
different  error factors to the overall error budget:
 
\begin{eqnarray}
\frac{\Delta\delta\ell}{b} &=&\Delta\alpha \,\sin\alpha_2\cos\delta_2+
\Delta\alpha_1\Big(\sin\alpha_2\cos\delta_2-
\sin\alpha_1\cos\delta_1\Big)+\nonumber\\[1.0mm]
&+&\Delta\delta \,\cos\alpha_2\sin\delta_2+
\Delta\delta_1\Big(\cos\alpha_2\sin\delta_2-
\cos\alpha_1\sin\delta_1\Big)+\nonumber\\[1.0mm] 
&+&\frac{\Delta b}{b}\,\Big(\cos\alpha_1\cos\delta_1-
\cos\alpha_2\cos\delta_2\Big)+\frac{\Delta\delta c_0}{b}+\nonumber\\[1.5mm]
&+&\Big[\frac{\Delta v}{c}\cos(\epsilon-\epsilon_0)
-\Delta\epsilon~\frac{v}{c} \sin(\epsilon-\epsilon_0)\Big]
\sqrt{(a^2+f^2)\cos^2(\psi-\psi_0)+k^2}-\nonumber\\[1.5mm]
&-&\Delta\psi~\frac{v}{c}\cos\epsilon
\sqrt{a^2+f^2}\sin(\psi-\psi_0),
\label{eq:del15c}
\end{eqnarray}

\noindent where we have used the following definitions for the situation
discussed:
\vskip 2pt
\begin{tabular}{rcl}
$\alpha=\alpha_2-\alpha_1$&:& {\tt RA} ~angle between the two stars 
\underline{\rm \em{inside}} the tile under consideration;\\
$\delta=\delta_2-\delta_1$&:& {\tt DEC} ~angle between the two stars 
\underline{\rm \em{inside}} the  tile under consideration.\\
&& [Note that the following identity is expected to hold for
these two \\
&&  angles $~\Rightarrow ~ \alpha^2+\delta^2\leq(\frac{\pi}{12})^2$~];\\
$\Delta\alpha $ &:& error in defining $\alpha$:
$~\Delta\alpha=\alpha_{\tt true}-\alpha_{\tt estimated}$;\\
$\Delta\delta $ &:& error in defining $\delta$:
$~\Delta\delta=\delta_{\tt true}-\delta_{\tt estimated}$;\\
$\Delta\alpha_1$&:& error in defining ~{\tt RA} angle, $\alpha_1$, for
                    the primary star\\
&& $\Delta\alpha_1={\alpha_1}_{\tt true}-{\alpha_1}_{\tt estimated}$;\\
 $\Delta\delta_1$&:&error in defining ~{\tt DEC} ~angle, $\delta_1$, for
                    the primary star\\
&& $\Delta\delta_1={\delta_1}_{\tt true}-{\delta_1}_{\tt estimated}$;\\
$\Delta b$&:&error in defining the baseline length $b$:
$~\Delta b=b_{\tt true}-b_{\tt estimated}$;\\
$\Delta \delta_{c_0}$ &:& error in defining the relative constant term for
the observations \\
&& \underline{\rm \em inside} the tile:  $~\Delta
\delta_{c_0}={\delta_{c_0}}_{\tt true}-
{\delta_{c_0}}_{\tt estimated}$;\\
$\Delta v, ~\Delta\epsilon, ~\Delta\psi$ &:& errors in knowledge of  
components of the three-dimensional spacecraft\\
&& velocity  vector, defined as usual ~$\Rightarrow ~\Delta$ =
{\tt true - estimated}.\\
\end{tabular}
\vskip 2pt 

Additionally,  the new quantities $\psi_0$ and $\epsilon_0$
are entirely defined by the coordinates of the two stars under
consideration  and are given as follows:
 
\begin{eqnarray}
\sin\psi_0&=&\frac{f}{\sqrt{a^2+f^2}}, ~\qquad~
\cos\psi_0=\frac{a}{\sqrt{a^2+f^2}}, \nonumber\\[1.5mm]
a&=&\cos^2\delta_2\cos^2\alpha_2 
-\cos^2\delta_1\cos^2\alpha_1, \nonumber\\[1.8mm]
f&=&\cos^2\delta_2\cos\alpha_2\sin\alpha_2-
\cos^2\delta_1\cos\alpha_1\sin\alpha_1.  
\label{eq:psi} \\[1.0mm]
\sin\epsilon_0&=&\frac{k}{\sqrt{(a^2+f^2)\cos^2(\psi-\psi_0)+k^2}}, 
\nonumber\\[1.0mm]
\cos\epsilon_0&=&\frac{\sqrt{a^2+f^2}\cos(\psi-\psi_0)}
{\sqrt{(a^2+f^2)\cos^2(\psi-\psi_0)+k^2}},\nonumber\\[1.8mm] 
k&=&\cos\delta_2\sin\delta_2\cos\alpha_2 
-\cos\delta_1\sin\delta_1\cos\alpha_1. 
\label{eq:epsilon}
\end{eqnarray}

 These
expressions (\ref{eq:del15c})-(\ref{eq:epsilon})  may now be used to study
propagation of astrometric errors to the first order. With these results one
may analyze the relativistic stellar aberration issue  analytically for a
different positions of the two stars of interest.  Note that in reality all
the quantities involved are both dynamical and stochastic functions of time. 
This fact we be used   in the Section \ref{sec:req} when we will
analyze the allocations adopted for  different constituents 
of the SIM astrometric error  budget.

\section{\Large Tile in the SIM nominal observing direction}
\label{sec:tile}
 
Analysis of the tolerable errors for the velocity aberration in a
general case is a complicated problem and should be attacked by using a
formal numerical treatment.  However, we may simplify the task by analyzing a
number of a  special cases which are expected to provide the most
stringent requirements on the velocity magnitude. Thus, one may expect that
the most driving requirement will come in the case when  the two stars are
aligned in a tile in  a such a way that the angular
separations between them is given by $\alpha=\frac{\pi}{12},~\delta=0$. It
turns out that this is indeed the case.  The compensation for  aberration
across the field of view of the interferometer in a direction parallel to
the baseline, will present the greatest challenge for SIM.   
To demonstrate this, we will discuss a more general
situation (that includes the case mentioned above), which allows one to obtain
an analytical expressions  useful for the future analysis.

\begin{figure}[t]
 \begin{center}
    \psfig{figure=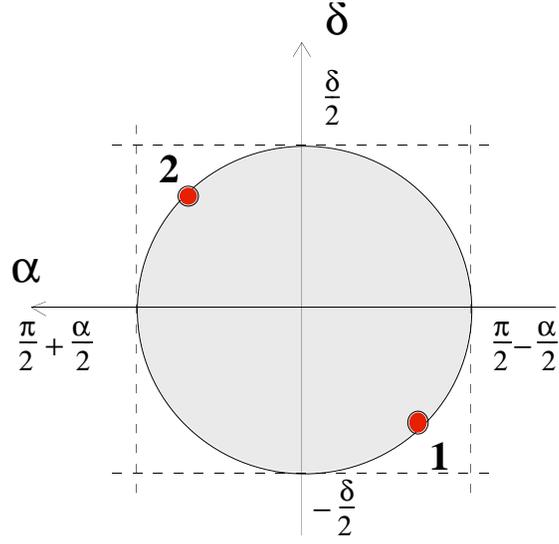,width=75mm,height=72mm}
     \caption{A tile in the SIM  nominal observing direction. Note that 
$\alpha^2+\delta^2\leq(\frac{\pi}{12})^2$. 
      \label{fig:sim_look}}
 \end{center}
\end{figure}

Let us define a tile in the SIM  nominal  observing  direction --- 
perpendicular to the baseline.  For the purposes of our analysis
we will  assume  that positions of the two stars  are symmetric with
respect to the point with coordinates
$(\alpha_0=\frac{\pi}{2},\delta_0=0)$.   Then the most suitable  
description for  the  nominal SIM observing mode is given by coordinates
of the primary and the secondary stars as follows:
{}
\begin{alignat}{2}
\alpha_1 &= ~~\,\frac{\pi}{2}-\frac{\alpha}{2},& ~~~\qquad
\alpha_2 &=\frac{\pi}{2}+\frac{\alpha}{2},\nonumber\\[2mm]
\delta_1 &=  -\frac{\delta}{2},& ~~~\qquad 
\delta_2 &=\frac{\delta}{2}.
\label{eq:tile}
\end{alignat} 
\noindent Now we can define a tile  for this   observing  direction 
as  a set of all the points for which the angular separations $\alpha$
and $\delta$ are  limited as: $\alpha^2+\delta^2\leq(\frac{\pi}{12})^2$. 
The resulted area constitutes the tile in  the  SIM nominal observing
direction and it is shown as a shaded area in the  Figure
\ref{fig:sim_look}. 


By substituting the coordinates for the two stars  Eq.(\ref{eq:tile}) in 
the expressions  (\ref{eq:psi})-(\ref{eq:epsilon}) we obtain the following
values for the constants involved:
 
\begin{eqnarray}
\begin{array}[c]{rcl} 
a&=&0,\nonumber\\[2.5mm]
f&=&-\sin\alpha \cos^2\frac{\delta}{2},\nonumber\\[2.5mm]
k&=&0,\nonumber\\[2.5mm]
a^2+f^2&=&\sin^2\alpha \cos^4\frac{\delta}{2}, \nonumber\\[2.5mm]
a^2+f^2+k^2&=&\sin^2\alpha \cos^4\frac{\delta}{2}.  
\end{array}
&~\Rightarrow ~&  
\begin{array}[c]{ccc}
 \begin{array}[c]{ccc} 
  \sin\psi_0&=&-1 \\[2.5mm]
  \cos\psi_0&=&0
 \end{array} 
\Bigg\} \quad  &\Rightarrow& ~~~\psi_0=-\frac{\pi}{2},\\
{}\\
 \begin{array}[c]{ccc} 
  \sin\epsilon_0&=&\,~0~\\[2.5mm]
  \cos\epsilon_0&=&\,~1~
 \end{array} 
\Bigg\} \quad  &\Rightarrow& ~~~\epsilon_0=0.
\end{array}
\label{eq:simsit}
\end{eqnarray}

\noindent Utilizing these results in the equation (\ref{eq:del15c}),
we obtain the expression necessary to analyze 
contributions of   different  error factors to the total  
differential OPD. This expression has the  following form: 

\begin{eqnarray}
\frac{\Delta\delta\ell}{b} 
&=&\Delta\alpha\,\cos\frac{\alpha}{2}\cos\frac{\delta}{2}~-~
\Delta\delta\,\sin\frac{\alpha}{2}\sin\frac{\delta}{2}~+~ 
\frac{2\Delta b}{b}\,\sin\frac{\alpha}{2}\cos\frac{\delta}{2}
~+~\frac{\Delta\delta c_0}{b}~-\nonumber\\[2mm]
&-&\Big[\Big(\frac{\Delta v}{c}\,\cos\epsilon
~-~\Delta\epsilon~\frac{v}{c}\,\sin\epsilon\Big)
\sin\psi~+~\Delta\psi~\frac{v}{c}\,\cos\epsilon\cos\psi\Big]
\sin\alpha\cos^2\frac{\delta}{2}. 
\label{eq:simmain2}
\end{eqnarray}

The obtained expression Eq.(\ref{eq:simmain2}) may now be used to analyze
the  propagation of errors in the future  astrometric observations with SIM. 

\subsection{Propagation of astrometric errors}
\label{sec:err}

In this section we will concentrate on obtaining the expressions that are
essential for derivations of the requirements on the quality of the spacecraft 
navigation data. These  requirements are imposed by the expected  accuracy of the
future  SIM astrometric measurements, such as determination of positions,
parallaxes and proper motions. 

In order to derive the necessary equations,  the first term in the expression 
(\ref{eq:simmain2}) $\Delta\delta\ell$ may equivalently be 
presented as $\Delta\delta\ell=\Delta(n\lambda_0)=
\Delta n\,\lambda_0+n\,\Delta\lambda_0$, where $\lambda_0$  is the operating
frequency and $n$ is an integer number. This term  vanishes because 
both $\lambda_0$ and $n$ are assumed to be known to a sufficiently 
high accuracy, thus $\Delta n=0, ~\Delta\lambda_0=0$. This is true
because SIM will be using it's fringe tracker to find a white light fringe
to perform the calibration of the instrument. Then the remaining
differentials
$\Delta v, ~\Delta\epsilon, ~\Delta\psi, ~\Delta\alpha,  ~\Delta\delta,
~\Delta b $ and
$\Delta\delta c_0$ will have to satisfy the  equation:   
{} 
\begin{eqnarray} 
\Delta\alpha\,\cos\frac{\alpha}{2}\cos\frac{\delta}{2}&-&
\Delta\delta\,\sin\frac{\alpha}{2}\sin\frac{\delta}{2}~=~
-\frac{\Delta\delta c_0}{b}-
\frac{2\Delta b}{b}\,\sin\frac{\alpha}{2}\cos\frac{\delta}{2}
+\nonumber\\[2mm]
+\Big[\Big(\frac{\Delta v}{c}\,\cos\epsilon
&-&\Delta\epsilon~\frac{v}{c}\,\sin\epsilon\Big)
\sin\psi ~+~ \Delta\psi~\frac{v}{c}\,\cos\epsilon\cos\psi\Big]
\sin\alpha\cos^2\frac{\delta}{2}.  
\label{eq:simmain0}
\end{eqnarray}
\noindent In order to simplify   further analysis we have separated the 
terms in the expression above in a such a way, so that the left side of this
equation represents the error in the measurement of absolute angular
separation  between the two stars on the sky, while the left side  shows the
main contributing factors to this quantity.

\subsubsection{Velocity knowledge requirement}

 One may expect that, in any given tile, the errors   $\Delta\alpha$ and
$\Delta\delta$ are normally distributed and 
uncorrelated.\footnote{Note that this is not true for a general case of
studying the stability of the  reference grid (\cite{Swartz98}). Thus, one
finds that the correlation  in {\tt RA} and {\tt DEC} becomes a source for
the zonal errors in the  analysis of the grid accuracy and stability.}  
Also, the errors due to the  orbital dynamics $\Delta v, ~\Delta\epsilon, 
~\Delta\psi$  at the chosen approximation may be treated as being
uncorrelated with the instrumental errors 
$\Delta b, ~\Delta c_0$.\footnote{A more general analysis should include a
possible correlation between the constant term $\Delta c_0$ and the two
angular components of velocity $\Delta\epsilon,  ~\Delta \psi$. We will
address this  possibility later in the Appendix \ref{appc}   and will
discuss the components ${\tt cov}(c_0,\epsilon)$ and ${\tt cov}(c_0,\psi)$
of the total covariance matrix.} Then, one obtains: {}
\begin{eqnarray} 
\sigma^2_\alpha\,\cos^2\frac{\alpha}{2}\cos^2\frac{\delta}{2}+
\sigma^2_\delta\,\sin^2\frac{\alpha}{2}\sin^2\frac{\delta}{2}
&=&\frac{\sigma^2_{\delta_{c_0}}}{b^2}+
\frac{4\sigma^2_b}{b^2}\,\sin^2\frac{\alpha}{2}\,\cos^2\frac{\delta}{2}
+\nonumber\\[2mm]
+\Big[\Big(\frac{\sigma^2_v}{c^2}\,\cos^2\epsilon
+\sigma^2_\epsilon~\frac{v^2}{c^2}\,\sin^2\epsilon\Big)
\sin^2\psi&+&\sigma^2_\psi~\frac{v^2}{c^2}\,\cos^2\epsilon\,\cos^2\psi\Big]
\sin^2\alpha\cos^4\frac{\delta}{2}.
\label{eq:simmain3a}
\end{eqnarray}

For the wide-angle astrometric observations with  {\small
FoR}$\sim\frac{\pi}{12}$, the  interferometer is expected to perform  at the
level of $\sigma_\alpha=4~\muas$ and $\sigma_\delta\sim 1~$mas.  Expression
(\ref{eq:simmain3a}) suggests that for observations in the direction
perpendicular to the baseline (e.q. when angle $\delta$ takes non-zero
values),  there will be   a  large contribution  coming from $\sigma_\delta$ to
the error budget, thus weakening the relativistic stellar aberration
requirement.    However, for the observations parallel to the baseline (e.q
$\delta=0$), influence of $\sigma_\delta$ is vanishes and one obtains the
following equation: 
{}
\begin{eqnarray} 
\sigma^2_\alpha\,\cos^2\frac{\alpha}{2}&=&
\frac{\sigma^2_{\delta_{c_0}}}{b^2}+
\frac{4\sigma^2_b}{b^2}\,\sin^2\frac{\alpha}{2}+\nonumber\\[2mm] 
&+&\Big[\Big(\frac{\sigma^2_v}{c^2}\,\cos^2\epsilon
+\sigma^2_\epsilon~\frac{v^2}{c^2}\,\sin^2\epsilon\Big)
\sin^2\psi+\sigma^2_\psi~\frac{v^2}{c^2}\,\cos^2\epsilon\,\cos^2\psi\Big]
\sin^2\alpha.
\label{eq:simmain1eq}
\end{eqnarray}

This expression depends on the orientation of the spacecraft velocity  
with respect to the SSB coordinate reference frame. Let us analyze the worst
case  of the  spacecraft orbital orientation.  This orientation is when
the spacecraft is moving towards/from the tile in the direction exactly
normal to the tile of interest. In this case, when
$\psi=\pm\frac{\pi}{2}, ~\epsilon=0$ (or in the direction   normal 
to the tile), from the equation Eq.(\ref{eq:simmain1eq})  one immediately
obtains expression that contains a requirement on the error in the 
magnitude of the spacecraft's  velocity:  
{}
\begin{eqnarray} 
\sigma^2_\alpha\,\cos^2\frac{\alpha}{2}&=&
\frac{\sigma^2_{\delta_{c_0}}}{b^2}+
\frac{4\sigma^2_b}{b^2}\,\sin^2\frac{\alpha}{2} +
\frac{\sigma^2_v}{c^2}\,\sin^2\alpha. 
\label{eq:simmain2a}
\end{eqnarray}

One   could also verify, that  errors in both sky angles 
$\psi$ and $\epsilon$   are related to that for the 
velocity magnitude $v$ and are given by   {}
\begin{equation}
\sigma^2_\psi=\sigma^2_\epsilon=\frac{\sigma^2_v}{v^2}.
\label{eq:sigma_angles}
\end{equation}

Note that a possible correlation between the constant term $\delta c_0$
and the two sky angles of velocity $\psi, \epsilon$ seriously affecting
these estimates (a more detailed analysis of this problem is given in
Appendix \ref{appc}). These errors, $\sigma_\epsilon$
and $\sigma_\psi$,   are  related to the  error in the velocity magnitude
$\sigma_v$ as follows:
\begin{eqnarray}
\sigma_\epsilon&=&\sqrt{\frac{\,\sigma^2_v}{v^2}+
\Big[\frac{\,\sigma_{\delta_{c_0}}}{b}
\frac{\,\rho(c_0,\epsilon)}{\,\sin\alpha}\,\frac{c}{v}\Big]^2}
-\frac{\,\sigma_{\delta_{c_0}}}{b}
\frac{\,\rho(c_0,\epsilon)}{\,\sin\alpha}\,\frac{c}{v}~\ge~0, \nonumber\\[2mm]
\sigma_\psi&=&\sqrt{\frac{\,\sigma^2_v}{v^2}+
\Big[\frac{\,\sigma_{\delta_{c_0}}}{b}
\frac{\,\rho(c_0,\psi)}{\,\sin\alpha}\,\frac{c}{v}\Big]^2}
-\frac{\,\sigma_{\delta_{c_0}}}{b}
\frac{\,\rho(c_0,\psi)}{\,\sin\alpha}\,\frac{c}{v}~\ge~0,
\label{eq:angles01}
\end{eqnarray}
\noindent where

\begin{tabular}{rl}
$\sigma_\epsilon, ~\sigma_\psi$ & --  the  errors in the two sky angles of the
 barycentric velocity vector $\vec{v}$;\hfill \\
$\rho(c_0,\epsilon), ~\rho(c_0,\psi)$ & --  correlation factors between
the constant term and the two sky angles\hfill \\
& ~~\,of the spacecraft's barycentric velocity,
$|\rho(c_0,\epsilon)|, |\rho(c_0,\psi)|\le 1$. 
\end{tabular} 

In addition to the velocity knowledge requirement one will have to impose two
additional constraints on the quality of  navigation data delivered by DSN.
These two constraints are on the sampling rate $\Delta t$ and positional
accuracy of the spacecraft determination $\Delta r$ together with the requirement
on the stochastic acceleration control during the time of astrometric
observations.

\subsubsection{Sampling rate and orbital position knowledge}

As we discussed previously, the  correction for the annual stellar aberration
may reach $\sim 20.5$ arcsec. This allows estimation of the time interval on
which this correction must be introduced in order to maintain the nominal
accuracy. Thus, contribution of only the velocity of the spacecraft to the
total error budget $\Delta\alpha$ may be obtained from Eqs.(\ref{eq:simmain0}).
For the motion in the plane of ecliptic with zero declination  
($\delta=\epsilon=0$), one will have:
\begin{eqnarray} 
\Delta\alpha\,\cos\frac{\alpha}{2} =
\Big(\frac{\Delta v}{c} \sin\psi ~+~
\Delta\psi~\frac{v}{c}\,\cos\psi\Big)\sin\alpha.
\label{eq:simm0}
\end{eqnarray}

How often one will needs to introduce a correction for the stellar aberration?
This question may be answered  directly by analyzing the motion 
of the spacecraft in the plane parallel to the tile,   when the  aberration
takes it's largest value. The latitude argument for circular motion
coincides with the mean anomaly, e.q.
$\psi=n\,t$, where
$n=\frac{2\pi}{P}$  and $P$ is the period of orbital motion of the
spacecraft. Taking this into account and setting ($\psi=0;\epsilon=0$), from  
equation (\ref{eq:simm0}) one obtains:
\begin{eqnarray} 
\Big(\frac{d\Delta\alpha}{d\psi}\Big)\,\cos\frac{\alpha}{2}\,\Delta\psi=
\frac{v}{c}\,\sin\alpha\,\frac{2\pi}{P}\Delta t,
\label{eq:simma1}
\end{eqnarray}
\noindent where  $\Delta\psi=n\,\Delta t$ and $\Delta t$  is the time
interval. After averaging, the last equation may be presented in a more useful
form (note,
$\sigma_{\alpha}\sim\overline{\frac{d\Delta\alpha}{d\psi}\,\Delta\psi}$), namely:
\begin{eqnarray} 
\sigma_{\alpha}\,\cos\frac{\alpha}{2}\ge
\frac{v}{c}\,\sin\alpha\,\frac{2\pi}{P}\Delta t,
\label{eq:simma2}
\end{eqnarray}

\noindent As a result, one will have to correct for the relativistic stellar
aberration with such a sampling rate $\Delta t$ that the expression on the
right-hand side of the equation above will be considerably less then on the
left side. 

The data sampling rate $\Delta t$ may be used to define the corresponding
accuracy of determination  of the  spacecraft's position  on it's orbit 
$\Delta r_{||}$. By decomposing  the spacecraft's orbital position vector 
on the component parallel to velocity vector and the one perpendicular   
as given by $\Delta \vec{r}=\Delta \vec{r}_{\perp}+\Delta \vec{r}_{||}$,
one may estimate the magnitudes for the both components involved.  The
expression for the magnitude of $\Delta \vec{r}_{||}$ may be obtained simply
as follows: 
\begin{eqnarray} 
\Delta r_{||}\le v_{\tt SIM} \,\Delta t.
\label{eq:simma23}
\end{eqnarray}

A requirement on the magnitude of the second component of the spacecraft
position vector $\Delta \vec{r}_{\perp}$ may be derived from another mission
goal, namely the expected accuracy for parallax determination. Thus, SIM is
expected to achieve a mission accuracy of the determination of parallax at
the level of $\sigma_\pi=1~\mu$as for a nearby stars. This goal may be
transformed into the requirement on the accuracy of the radial  component of
the barycentric spacecraft position $\Delta \vec{r}_{\perp}$. On may show
that the  error, introduced by annual parallax into the 
astrometric observations of a pair of stars separated on the sky by  
angle $\alpha$   (with $\delta=0$), may be  given as below:
\begin{equation}
\delta \alpha=\alpha_1-\alpha_2=\frac{r^{[\pi]}_\perp}{\cal D}2\cos
\Big(\psi-\frac{\alpha_1+\alpha_2}{2}\Big)\sin\frac{\alpha_2-\alpha_1}{2},
\end{equation}
\noindent where ${r^{[\pi]}_\perp}$ is the spacecraft barycentric distance
(the  superscript $[\pi]$ is used to separate parallactic distance 
requirement from those derived with the help of some other methods),
and ${\cal D}$ is the distance to the object of study. Then for the tile in
the SIM nominal observing direction, defined by Eqs.(\ref{eq:tile}), and for
the motion perpendicular to the tile ($\psi=\pm\frac{\pi}{2}$) one may
obtain  following relation:
\begin{eqnarray} 
\Delta r^{[\pi]}_\perp\le  \, \sigma_\pi\,{\cal
D}\frac{1}{2\sin\frac{\alpha}{2}}.
\label{eq:simma235}
\end{eqnarray}
\noindent  This relation suggests that the closer the
distance
${\cal D}$ to the observed object the better should be the knowledge of the
barycentric position of the spacecraft. 

Accuracy of the spacecraft orbital position determination $\Delta r$ is
one of the navigational products that will be provided by means of DSN.
However, one should not expect that DSN will provide such a continuous
trucking of the spacecraft with the  data sampling rate $\Delta t$. This
constraint (primarily cost of such an extended DSN commitment) implies that a
real-time on-board correction must be done in order to provide the conditions
necessary for astrometric observations with expected accuracy. Note that this
problem may be already solved by means of the SIM
multiple-baseline architecture, when the two guide interferometers will
take care of a number of similar effects. In any case, this question is out
of scope of the present analysis, but it should be  re-visited as a subject
for separate study.

\subsubsection{Acceleration knowledge constraint}

The two additional constraints, namely on the quality of the DSN data  
indirectly imposed by the expected  accuracy for  proper motions determined for
a number of stars during the mission's five-years live time, which will be  of
the order of
$\mu_\alpha=1~\mu$as/yr. Such a constraint may be also obtained  from the
Eq.(\ref{eq:simm0}). By differentiating this equation with respect to time one
obtains:
\begin{eqnarray} 
\frac{d \Delta\alpha}{d t}\,\cos\frac{\alpha}{2} =
\Big(\frac{1}{c}\frac{d\Delta v}{d t} \sin\psi ~+~
\frac{d\Delta\psi}{d t}~\frac{v}{c}\,\cos\psi\Big)\sin\alpha.
\label{eq:simma3a}
\end{eqnarray}
\noindent The time derivatives on the right-hand side of this
equations are the components of the spacecraft acceleration. Indeed, one may
decompose vector of this acceleration onto the components normal to the
orbit (radial) and the tangential one  as follows
$\vec{a}=\vec{a}_\perp+\vec{a}_{||}$, where
$a_{\perp}=\frac{\overline{d\Delta v}}{d t}$ and
$a_{||}=\frac{\overline{d\Delta \psi}}{d t}$. Assuming that the errors in both
accelerations are normally distributed and uncorrelated, this equation may be
averaged and presented in the following useful form:
\begin{equation}
\sigma^2_{\mu_\alpha}\,\cos^2\frac{\alpha}{2} =
\Big[\frac{1}{c^2}\Big(\frac{\overline{d\Delta v}}{d
t}\Big)^2\sin^2\psi ~+~
\Big(\frac{\overline{d\Delta \psi}}{d t}\Big)^2~\frac{v^2}{c^2}\,\cos^2
\psi\Big]\sin^2\alpha.
\label{eq:simma33b}
\end{equation}

For the motion of the spacecraft in the direction normal to the tile of
interest $(\psi=\pm\frac{\pi}{2})$, one may obtain  expression 
that imposes a requirements on the  knowledge of the spacecraft
radial acceleration  the following form:
\begin{eqnarray} 
\sigma_{\mu_\alpha}\,\cos\frac{\alpha}{2} \ge
 \frac{1}{c}\frac{\overline{d\Delta v}}{d t} \sin\alpha.
\label{eq:simm321}
\end{eqnarray} Similarly, for the motion of the spacecraft in the
direction parallel to the tile  $(\psi=0)$, one  will have a  
requirement on the  knowledge of the angular acceleration:

\begin{eqnarray} 
\sigma_{\mu_\alpha}\,\cos\frac{\alpha}{2} \ge
 \frac{v}{c}\frac{\overline{d\Delta \psi}}{d t} \sin\alpha.
\label{eq:simm322}
\end{eqnarray} 

These two  equations suggest that, for accurate determination of  proper
motions one will need to control all the non-gravitational forces acting on the
spacecraft at a very high level of accuracy. Note, for the expected orbital
parameters of SIM, the control of the radial accelerations  should be at the
level of 
$\sim 1.2\times 10^{-13}$ km/s$^2$. 

\section{\Large Relativistic stellar aberration requirements}
\label{sec:req}

In this Section we will present the estimates for a different strategies of
accessing the influence of the relativistic stellar aberration  on the
expected accuracy of astrometric measurements with SIM.

We will begin from the expressions describing the uncorrelated data
Eqs.(\ref{eq:simmain2a})-(\ref{eq:sigma_angles}). 
Remembering that angles
$\alpha,\delta$  vary in the range given by
$\alpha^2+\delta^2\leq(\frac{\pi}{12})^2$ for which $\cos\frac{\alpha}{2}$  
never vanishes, one can divide the both sides of the equation
(\ref{eq:simmain2a}) on $\cos^2\frac{\alpha}{2}$:
{}
\begin{eqnarray} 
\sigma^2_\alpha &=&
\frac{\sigma^2_{\delta_{c_0}}}{b^2 \cos^2\frac{\alpha}{2}}+
\frac{4\sigma^2_b}{b^2}\,\tan^2\frac{\alpha}{2} +
\frac{4\sigma^2_v}{c^2}\,\sin^2\frac{\alpha}{2}.
\label{eq:simmain3}
\end{eqnarray}  

\noindent The obtained equation represents ellipsoid  with  
half-axis  defined as follows:\footnote{A more general expression for the
stellar aberration only may be obtained from 
Eqs.(\ref{eq:del15c})-(\ref{eq:epsilon}). It depends on the coordinates of
both stars under consideration and has the following from:
$\sigma_{\alpha_v}=\frac{\sigma_v}{c}\,\frac{\sin(\alpha_2-\alpha_1)}
{\sin\alpha_2}$.}  {}
\begin{eqnarray} 
\sigma_{\delta_{c_0}}&=&\sigma_\alpha \,b\,\cos\frac{\alpha}{2},
\label{eq:simmain3c1}\\
 \sigma_b &=&\sigma_\alpha \,\frac{b}{2}\,\cot\frac{\alpha}{2}, 
\label{eq:simmain3c2} \\
\sigma_v  &=&\sigma_\alpha \,\frac{c}{2 \sin\frac{\alpha}{2}}.  
 \label{eq:simmain3c3}
\end{eqnarray} 
The necessary expressions for the tolerable  errors in  
the components of the three-dimensional  vector of the spacecraft's 
velocity  may be obtained directly from
Eqs.(\ref{eq:sigma_angles}),(\ref{eq:simmain3c3}) as 

\begin{equation}
\sigma_\psi=\sigma_\epsilon = \sigma_\alpha
\,\frac{c}{v_{\tt SIM}}\,\frac{1}{2 \sin\frac{\alpha}{2}}.
\end{equation}

In addition to these four constraints we will have to define the other three 
discussed in the previous Section \ref{sec:err}, namely 1) the data sampling
rate, $\Delta t$; 2) the accuracy of the orbital position determination, $\Delta
r_{\perp}$ and $\Delta r_{||}$;  and 3)  the accuracy of compensation of the
non-gravitational accelerations, $a_\perp=\frac{\overline{d\Delta v}}{d t}$
and $a_{||}=\frac{\overline{d\Delta \psi}}{d t}$. By utilizing equations
(\ref{eq:simma2})-(\ref{eq:simma235}) together with  
Eqs.(\ref{eq:simm321}),(\ref{eq:simm322}),  these constraints may be presented by
the  following set of equations:
\begin{eqnarray} 
\Delta t &\le&
\sigma_\alpha\,\frac{c}{2\sin\frac{\alpha}{2}}\,\frac{P}{2\pi v_{\tt SIM}},
\label{eq:simc31}\\ 
\Delta r_{||} \leq \sigma_\alpha
\,\frac{c}{2 \sin\frac{\alpha}{2}}\,\frac{P}{2\pi},&&
\Delta r^{[\pi]}_{\perp} \leq \, \sigma_\pi\,{\cal D}\,\frac{1}{2
\sin\frac{\alpha}{2}},
\label{eq:simm32}\\
\frac{\overline{d\Delta v}}{d t} \le
\sigma_{\mu_\alpha}\,\frac{c}{2 \sin\frac{\alpha}{2}}, &&
\frac{\overline{d\Delta \psi}}{d t} \le
\sigma_{\mu_\alpha}\,\,\frac{c}{v_{\tt SIM}}\,\frac{1}{2
\sin\frac{\alpha}{2}}.
\label{eq:simc3b}
\end{eqnarray}

\noindent These  five requirements are the derived constraints on the
orbital motion of the spacecraft and the data sampling rate. While the
stellar aberration constraint $\sigma_v$ is the most important among
those imposed by the orbital dynamics, we think that it is useful to have 
all the requirements related to the motion of the spacecraft spelled
out. 

Let us assume that a different error-budget constituents are contributing
individually to the total error budget with an individual astrometric error
$\sigma_\alpha = \Delta k ~~\mu$as, where  $\Delta k$ is some number.  By taking
into account the numeric   values for  different quantities involved   {}
\begin{alignat}{2}
\sigma_\alpha  = & ~\Delta k ~~\mu{\rm  as}, ~~&~~ \qquad  
c~=&~2.997292\times 10^{11} ~{\rm  mm/s},\nonumber \\ 
b =&~10.50 ~{\rm  m}, &~~ \qquad 1~\mu{\rm  as} ~=&
~4.84814\times 10^{-12} ~{\rm  rad},\nonumber \\   
\sigma_{\mu_\alpha} =& ~\Delta l ~~\mu{\rm  as/yr}, ~~&~~ \qquad  
\sigma_\pi~=&~~\Delta  q ~~\mu{\rm  as},\nonumber \\
v_{\tt SIM}\approx v_\oplus ~= &~2.98 \times 10^7~~{\rm  mm/s}, ~&~
\qquad   P~=&~3.1536\times 10^{7} ~{\rm  s} 
\label{eq:const}
\end{alignat}

\noindent one could compute tolerable astrometric contributions introduced
by the different errors budget constituents. The largest effect (the worst
case) will be in the case of maximal separation between the two stars
which will equal to the field of regard of the instrument 
$\alpha_{\tt max} = ${\small FoR}$ \equiv\frac{\pi}{12}$. 
 Substituting the numerical quantities from (\ref{eq:const}) we immediately
obtain the following values for various  components of the error budget:
{}
\begin{eqnarray} 
\sigma_{\delta_{c_0}}&=& ~\,50.470 ~\Delta k ~~~{\rm  pm},\nonumber\\
 \sigma_b &=& 193.332 ~\Delta k ~~~{\rm  pm}, \nonumber\\
\sigma_v  &=& ~~~5.566~\Delta k~~~{\rm  mm/s},\nonumber\\  
\sigma_\psi=\sigma_\epsilon &=& ~\,38.529 ~\Delta k~~~{\rm  mas};
 \label{eq:simmain4}
\\[11pt]
\Delta t  &=& ~~\,0.938 ~\Delta k ~~~\,{\rm  s},\nonumber\\
\Delta r_{||}= ~27.952 ~\,\Delta k ~~~{\rm  km},  ~~~~~~~&&~~~\Delta
r^{[\pi]}_\perp = ~573 ~\,\Delta q \,\Big(\frac{\cal D}{1 ~{\rm 
pc}}\Big)~~{\rm  km},
\nonumber\\
\frac{\overline{d\Delta v}}{d t} = ~~\,11.132~\Delta l~~~~ {\rm mm/s/yr},&&
~~\frac{\overline{d\Delta \psi}}{d t}= ~~77.058~\Delta l~~~~{\rm  mas/yr}, 
 \label{eq:simmain42}
\end{eqnarray}  
 
\noindent where we have assumed  that  the barycentric velocity of the 
spacecraft is  approximately equal to that of Earth' orbital motion around 
the sun, (e.g. $c/v_{\tt SIM} = 10058$), thus  giving us the requirement for
the two sky angles of the velocity vector.

The constraints on the spacecraft's position are given by the
Eqs.(\ref{eq:simmain42}). It is worth noting that the parallactic
requirement on the spacecraft barycentric position $\Delta r^{[\pi]}_\perp$ is
easy to meet for a distant objects, because it grows linearly with distance.
[Note that for the observations of the solar system objects one will have to use
a completely different observational strategy.] The accuracies of the DSN
navigation are much superior than it needed to satisfy this parallactic
requirement. Note that the general relativistic deflection of light is also
distance-dependent effect. One may think that this dependence may produce an
independent constraint on
$\Delta r^{[\pi]}_\perp$. However, a crude  estimate shows that this is not the
case and general relativity does not require a significant  accuracy of
knowledge of the barycentric distance of the spacecraft.  This is why we will
omit the constraint on $\Delta r^{[\pi]}_\perp$ from our studies. The last two
constraints in Eqs.(\ref{eq:simmain42}), on the two components of the
spacecraft  acceleration. These constraints are suggesting that the total
uncompensated error in the acceleration of the craft over a half-orbit 
(half-year time interval) of the mission should not exceed these numbers. 
The first requirement may  be re-written in  more  familiar units as 
$\frac{\overline{d\Delta v}}{d t} = 
\,3.53\times 10^{-13}~\Delta l~~{\rm  km/s^2},$ which again implies
quite a significant navigational involvement or a real-time on-board processing. 

\begin{figure}[t]
 \begin{center}
\rotatebox{90}{\hskip 68pt  Fringe intensity }
\hskip -60pt
    \psfig{figure=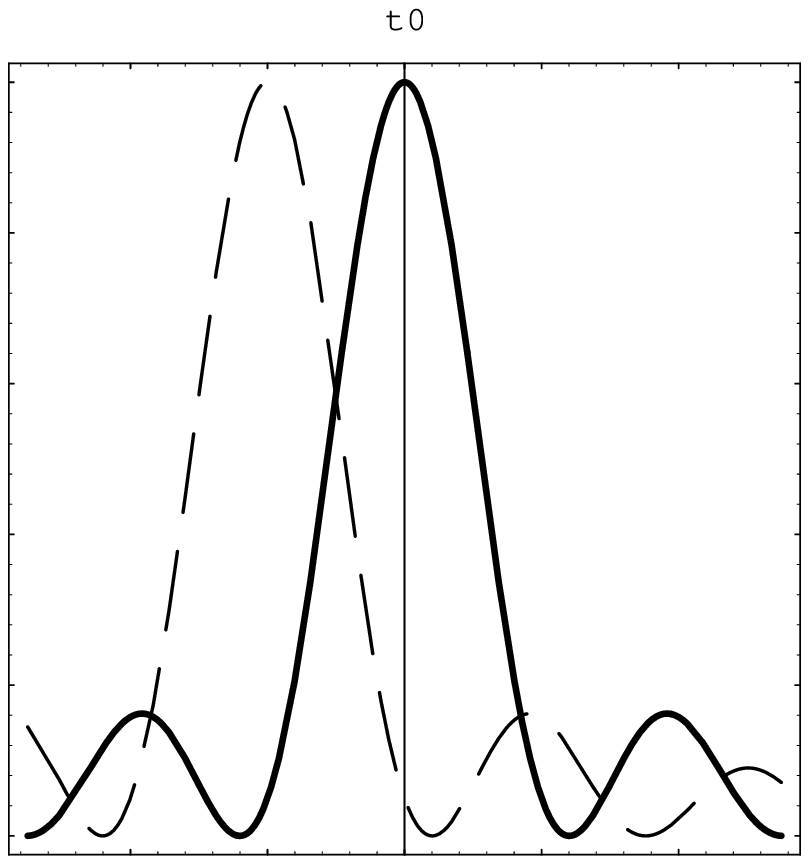,width=90mm,height=80mm}

\rotatebox{0}{\hskip 18pt Spatial fringe resolution drift}
     \caption{{Spatial fringe drift introduced by the
relativistic stellar aberration.  Suppose the data are taken with the
sampling rate $\Delta t$, then the thick curve -- is the  fringe intensity at the
beginning of observations at time $t_0$, while the dotted curve is plotted for
the position of the same  fringe after time interval $\Delta t$ has elapsed.}
      \label{fig:drift}}
 \end{center}
\end{figure}

The  worst case observation scenario will be realized for those angles $\psi$
and $\epsilon$ which will maximize the corresponding multipliers in formulae
(\ref{eq:simmain1eq}). Thus, the motion of the spacecraft in the direction
perpendicular to the  tile, or ~$(\psi\rightarrow\psi_\perp=-\frac{\pi}{2},
~\epsilon=0)$, will provide  the most stringent requirement  for  the
accuracy of knowledge of the magnitude of barycentric  velocity  of the
spacecraft. 
Additionally, the motion in the direction parallel to the  tile,
or ~$(\psi\rightarrow\psi_{||}=0,\pi; ~\epsilon=0)$, will provide  the
most stringent requirement  for  the accuracy of knowledge of the angular
position of the spacecraft with respect to the solar system barycenter
and, therefore, it introduces constraint on the data sampling rate $\Delta t$.
This   rate  is  necessary to apply for on-board navigation
in order to correct for the stellar aberration. The nature of this
correction for a single-baseline interferometer is presented by the Figure
\ref{fig:drift}.


Relativistic stellar aberration introduces a spatial fringe drift due to
the motion of the craft. This spatial fringe drift may be corrected if one
will be sampling data with the rate $\Delta t$. This time interval tells how
often one will need to introduce a correction to the velocity  of a
single-baseline interferometer in order to provide the necessary dynamical
conditions for astrometric measurements with accuracy  
$\sigma_\alpha$ (i.e. for the starts separated by FoR=$\alpha$). SIM is the
multiple baseline interferometer. The main function of the two guide
interferometers is to provide this sort of anchoring to the sky by  providing
a set of  differential navigation parameters.  Moreover, for the direction of
motion parallel to the tile of interest, the accuracy of the angular position
of the spacecraft is more important then it's velocity magnitude, thus 
relaxing quite a bit the corresponding values for $\Delta k$ in 
Eqs.(\ref{eq:simmain42}).  So, to the first order,  the requirement imposed 
by $\Delta t$ does not  significantly  impacting  SIM astrometric campaign,
but it is important enough to consider it for further studies.

\subsection{Two ways of direct establishing the error allocations}

In this Section we will present  analysis of different methods for
establishing the errors allocations for the  stellar aberration. The expressions
Eq.(\ref{eq:simmain4}) are representing the set of  requirements (including the
stellar aberration) for the astrometric observations with accuracy of
$\sigma_\alpha=\Delta k~\mu$as in the direction parallel to baseline.  By
varying the coefficient $\Delta k$ we will generate the  sets of requirements
for $\sigma_{\delta_{c_0}}, \sigma_b, \sigma_v, \sigma_\psi,$  and 
$\sigma_\epsilon$ that correspond to these different methods of error
allocations. 

There exist a number of different ways how to estimate contributions of
different constituents to the total astrometric error budget.
For a more accurate analysis,  one should perform a numerical simulations in
order to estimate the impact of those constituents on the astrometric grid
performance. At this point  we would like to analyze a number of possible
limiting cases, in order to provide a taste of different flavors to this
problem. To do this we will use the results presented by the set
Eqs.(\ref{eq:simmain4}).  In particular, we will analyze the constraints
that should be placed on different error budget constituents in the cases
when one choose to make the assumptions based on either single measurement 
accuracy or the expected wide-angle mission accuracy.

\subsubsection{Single observation accuracy requirements}

SIM is being designed to be able to achieve astrometric
accuracy of a single measurements of the order of $\sigma_\alpha=8~\mu$as. This
is  first useful number to use, in order to derive the maximal stellar
aberration error, tolerable during the tile observation.  
The corresponding estimate may be obtained by substituting the
value  $\Delta k =8$ into Eqs.(\ref{eq:simmain3c1})-(\ref{eq:simmain3c3}). 
As a result, we obtain the following values for  different 
terms in the equation  Eq.(\ref{eq:simmain3}):  
\begin{eqnarray} 
\sigma^{[8]}_{\delta_{c_0}} &=& ~\,403.76   ~~~{\rm  pm},\nonumber\\
 \sigma^{[8]}_b &=& 1546.66   ~~~{\rm  pm}, \nonumber\\
\sigma^{[8]}_v  &=& ~~~44.53 ~~~{\rm  mm/s},\nonumber\\  
\sigma^{[8]}_\psi=\sigma^{[8]}_\epsilon &=& ~\,308.23  ~~~{\rm  mas};
 \label{eq:simmain5} \\[11pt]
\Delta t^{[8]}  &=& ~~~~\,7.50 ~~~{\rm  s},\nonumber\\
\Delta r^{[8]}_{||} &=& ~\,223.62  ~~~{\rm  km}, \nonumber\\
\frac{\overline{d\Delta v}}{d t}^{[\Delta l=8]} = ~~\,44.53 ~~~{\rm 
mm/s/yr}, &&
~~\frac{\overline{d\Delta \psi}}{d t}^{[\Delta l= 8]}
= ~~616.46~~~{\rm  mas/yr} 
 \label{eq:simmain52}
\end{eqnarray}  

\noindent A reasonable improvement, for the numbers presented above, comes
from the statement that a contribution of any component of the total error
budget in the right-hand side of the equation  (\ref{eq:simmain3}) should not
exceed  10\% of the total variance  a single accuracy of
$\sigma^2_\alpha$. This gives  a different correction factor $\Delta k^{0.1}$
which is calculated to be $\Delta k^{0.1}=8\sqrt{0.1}=2.52982$. The resulting
numbers for the contributions to the error budget in a single
measurement mode are follows:  
\begin{eqnarray} 
\sigma^{[2.5]}_{\delta_{c_0}}&=& ~\,127.68   ~~~{\rm  pm},\nonumber\\
 \sigma^{[2.5]}_b &=& ~\,489.10   ~~~{\rm  pm}, \nonumber\\
\sigma^{[2.5]}_v  &=& ~~~14.08 ~~~{\rm  mm/s},\nonumber\\  
\sigma^{[2.5]}_\psi=\sigma^{[2.5]}_\epsilon &=& ~~~97.47  ~~~{\rm  mas}; 
 \label{eq:simmain5a}\\[11pt]
\Delta t^{[2.5]}  &=& ~~~~\,2.37 ~~~{\rm  s},\nonumber\\
\Delta r^{[2.5]}_{||} &=& ~~~70.72  ~~~{\rm  km}, \nonumber\\
\frac{\overline{d\Delta v}}{d t}^{[\Delta l=2.5]} = ~~~14.08 ~~~{\rm 
mm/s/yr}, &&
~~\frac{\overline{d\Delta \psi}}{d t}^{[\Delta l= 2.5]}
= ~~194.94~~~{\rm  mas/yr} 
 \label{eq:simmain5a2}
\end{eqnarray} 

It is important to point out  that the velocity aberration issue
has two flavors to it, in a sense that it is influencing both  the narrow
angle observations and (by the procedure of the astrometric grid
reduction)  the wide angle ones.  This mean that the error coming from the
velocity aberration knowledge inside one single tile, will propagate to the
total accuracy of the   wide angle observations. 
This is why it is important to estimate the corresponding maximal 
tolerable errors based on the  expected wide angle astrometric accuracy.

The set of requirements represents
Eqs.(\ref{eq:simmain5a}),(\ref{eq:simmain5a2}) an optimistic expectation on the
accuracy  of the velocity determination. Due to the reason that the astrometric
accuracy will be improving as mission progresses,  this set of relativistic
stellar aberration requirements will be easily met by the DSN navigation.
However, the assumption based on the single measurement accuracy is
over-optimistic and we need to consider the most driving cases presented by the
expected wide-angle mission accuracy.

\subsubsection{Mission accuracy requirements}
\label{sec:mission}

The wide angle astrometric observations with SIM are
expected to be with a mission accuracy of $\sigma_\alpha=4~\mu$as. 
Using this number,  one may derive a different set of requirements for SIM,
which will be exactly twice smaller than the ones given by
Eq.(\ref{eq:simmain5}). Additionally, by applying a conservative 10\% approach
(described above), one may derive another set of  estimates. The wide angle
astrometric mode  together with the conservative
approach,  gives for $\Delta k$ the  number $\Delta
k^{0.1}=4\sqrt{0.1}=1.26491$.  This value of the correction factor results in
the following  contributions to the error budget:  
\begin{eqnarray} 
\sigma^{[1.3]}_{\delta_{c_0}}&=& ~\,63.84  ~~~{\rm 
pm},\label{eq:simmain51}\\
 \sigma^{[1.3]}_b &=& 244.55   ~~~{\rm  pm}, \label{eq:simmain52a}\\
\sigma^{[1.3]}_v  &=& ~~~7.04 ~~~{\rm  mm/s},\label{eq:simmain53}\\  
\sigma^{[1.3]}_\psi=\sigma^{[1.3]}_\epsilon &=& ~\,48.74  ~~~{\rm  mas}; 
 \label{eq:simmain54}\\[11pt]
\Delta t^{[1.3]}  &=& ~~~1.19 ~~~{\rm  s},\label{eq:simmain531}\\
\Delta r^{[1.3]}_{||} &=& ~\,35.36  ~~~{\rm  km}, \label{eq:simmain532}\\
\frac{\overline{d\Delta v}}{d t}^{[\Delta l=1.3]} = ~~~7.04 ~~~{\rm 
mm/s/yr}, &&
~~\frac{\overline{d\Delta \psi}}{d t}^{[\Delta l= 1.3]}
= ~97.47~~~{\rm  mas/yr}. 
 \label{eq:simmain542}
\end{eqnarray} 

There is another important issue which hasn't been addressed yet. That is
a possible correlation between the calibration term $\delta_{c_0}$ and the
velocity sky-angles $(\psi,\epsilon)$ given by
Eq.(\ref{eq:angles01}).   To estimate the influence of this possible
correlation, we assume  that the two correlation coefficients for the two
angles involved are equal $\rho(c_0,\psi)=\rho(c_0,\epsilon)=\rho_0$. Then, by
using the   expressions  (\ref{eq:angles01}) together with the estimates
Eqs.(\ref{eq:simmain3c1})-(\ref{eq:simmain3c3}) one could obtain an
improved expression in order  to derive  requirements on the  accuracy of
knowledge of the two sky-angles of the spacecraft's velocity:

\begin{equation}
\sigma_\epsilon= \sigma_\psi
=\sigma_\alpha\,\frac{c}{v}\,\frac{1}{2\sin\frac{\alpha}{2}}
\Big(\sqrt{1+\rho^2_0}-\rho_0\Big)=\sigma_v 
\Big(\sqrt{1+\rho^2_0}-\rho_0\Big)\ge0.
\label{eq:corr}
\end{equation}
\noindent It is important to point out, that for the worst case of highly
correlated quantities ($\psi,\epsilon$) and $c_0$, e.q. $\rho_0\sim1$ the result
Eq.(\ref{eq:simmain54}) will have to be further reduced by a
factor of $\sqrt{2}-1=0.4142$, thus tightening the  requirements for the
accuracy of knowledge of the  two velocity sky-angles  as $\sigma_\epsilon
= \sigma_\psi\sim20$ mas. [Note that this estimate is for the worst case of
correlation, when $\rho \sim 1$. For the case, when $\rho_0 \sim 0.5$
this correction factor is almost twice relaxed, namely
$(1-0.5^2)^\frac{1}{2}-0.5^2=0.868$.]  This example suggests that a
possible correlation between the velocity components and the constant term
in the narrow-angle observations may put an additional demand on the quality
of the DSN data. Thus for the worst case scenario, DSN will have to
deliver   data for all three components of the velocity vector with
considerably smaller errors, say   {}
\begin{eqnarray} 
\sigma^{\tt corr}_v  &=& ~~~2.916 ~~~{\rm  mm/s},\label{eq:simmain71}\\  
\sigma^{\tt corr}_\psi = \,\sigma^{\tt corr}_\epsilon &=& 
~\,20.186  ~~~{\rm  mas}. 
 \label{eq:simmain72}
\end{eqnarray} 

The two numbers above are the very pessimistic estimates for the required
accuracy of determination the spacecraft velocity vector and are given for
the worst case of  highly correlated data. We have obtained this
requirement by using  the number for the expected mission accuracy for
wide-angle astrometric observations together with a conservative
allocation\footnote{This estimate was derived from the statement that the
contribution of any component of the error budget in the right side of the
equation  (\ref{eq:simmain3}) should not exceed  10\% of the total variance
$\sigma^2_\alpha$, with $\sigma_\alpha=4~\mu$as.} for different errors in the
total astrometric error budget.

\subsection{Currently adopted error allocations}

Presently the analysis of not only the stellar aberration, but also a number of
the other important issues  is complicated by the fact that a
realistic model for the spacecraft and the instrument is absent. This is
why the currently adopted values in the error budget were chosen more or
less intuitively. Moreover, these numbers are the same as they were for  the
Earth-orbiting mission study. Now, when SIM is designed to be placed on  the
Earth trailing SIRTF-like solar orbit, these numbers should be verified for a
better justification. The reason for doing that is not only the need for
justification of the number of  mm/s (this number is the same for both orbits),
but rather  time and  the level of DSN commitment, necessary to reach
the desired   accuracy of velocity determination.

 Remember that the numbers in Eqs.(\ref{eq:simmain54}) were obtained
based on the assumption that  the errors on the right-hand side of the
Eq.(\ref{eq:simmain2a}) are  forming an  ellipsoid with half-axes given by
Eqs.(\ref{eq:simmain3c1})-(\ref{eq:simmain3c3}). In reality, one will have
to minimize each constituent of the total error budget in a such a way that
in any given time the sum of the terms on the right-hand side  of the
equation (\ref{eq:simmain2a}) will not be larger then the expected variance
$\sigma^2_\alpha$.   Currently, the error budget estimations allocates
for the stellar aberration $\sigma_{\theta_v}=36$ pm. The corresponding   
$\Delta k$ is then estimated to   be of order   $\Delta k=0.701$
[based on $\sigma_{\theta_v}\equiv(\sigma^2_{\alpha_v}+
\sigma^2_{\delta_v})^\frac{1}{2}
\approx\sigma_{\alpha_v}=b\frac{\sigma_v}{c}=36$ pm and 
$b$=10.50~m]. The requirements  now will have to be modified as follows:
\begin{eqnarray} 
\sigma^{[0.7]}_{\delta_{c_0}}&=& ~\,35.38   ~~~{\rm  pm},
\label{eq:cc}\\
 \sigma^{[0.7]}_b &=& 135.53   ~~~{\rm  pm}, \\
\sigma^{[0.7]}_v  &=& ~~~3.90 ~~~{\rm  mm/s},
\label{eq:simvel3}\\  
\sigma^{[0.7]}_\psi=\sigma^{[0.7]}_\epsilon &=& ~\,27.01  ~~~{\rm  mas};
 \label{eq:simmain6}\\[11pt]
\Delta t^{[0.7]}_{||}  &=& ~~~0.66 ~~~{\rm  s},\label{eq:simmain61}\\
\Delta r^{[0.7]} &=& ~\,19.59  ~~~{\rm  km}, \label{eq:simmain62}\\
\frac{\overline{d\Delta v}}{d t}^{[\Delta l=0.7]} =~~~3.90 ~~~{\rm 
mm/s/yr}, &&
~~\frac{\overline{d\Delta \psi}}{d t}^{[\Delta l= 0.7]}
= ~~54.02~~~{\rm  mas/yr} 
 \label{eq:simmain642}
\end{eqnarray}  

Despite the fact that we have a reasonable gap between our estimates
Eq.(\ref{eq:simmain53}) and the best experimental guess given by
Eq.(\ref{eq:simvel3}), there some other factors that are necessary to
consider. As we demonstrated previously, a possible correlation between the
constant term and the two velocity sky-angles Eqn.(\ref{eq:corr}) may
completely eliminate this gap and further reduce the estimates
presented  in the Section \ref{sec:mission} (for the worst case of highly
correlated data). This fact   minimizes a tolerable errors
Eqs.(\ref{eq:simmain51})-(\ref{eq:simmain542}), reducing those down to the
values Eqs.(\ref{eq:cc})-(\ref{eq:simmain642}).  

Accounting for this correlation directly in the expressions above, leading
to $\sigma_v =1.616$  mm/s. However, remember that the contribution of
stellar aberration to the  total error budget was chosen to be
$\sigma_{\theta_v}=36$ pm.  In order to see whether or not this number
should be kept as a maximum tolerable  stellar aberration error,   this
problem  should be addressed with a formal numerical treatment. In the mean
time, we recommend that $\sigma_{\theta_v}$  be reduced to, say
$\sim25$ pm. 


\begin{figure}[ht]
 \begin{center}
\rotatebox{90}{\hskip 16pt  Maximal ~error in velocity,  
~$\sigma_v$, ~[mm/s]}
\hskip -50pt
    \psfig{figure=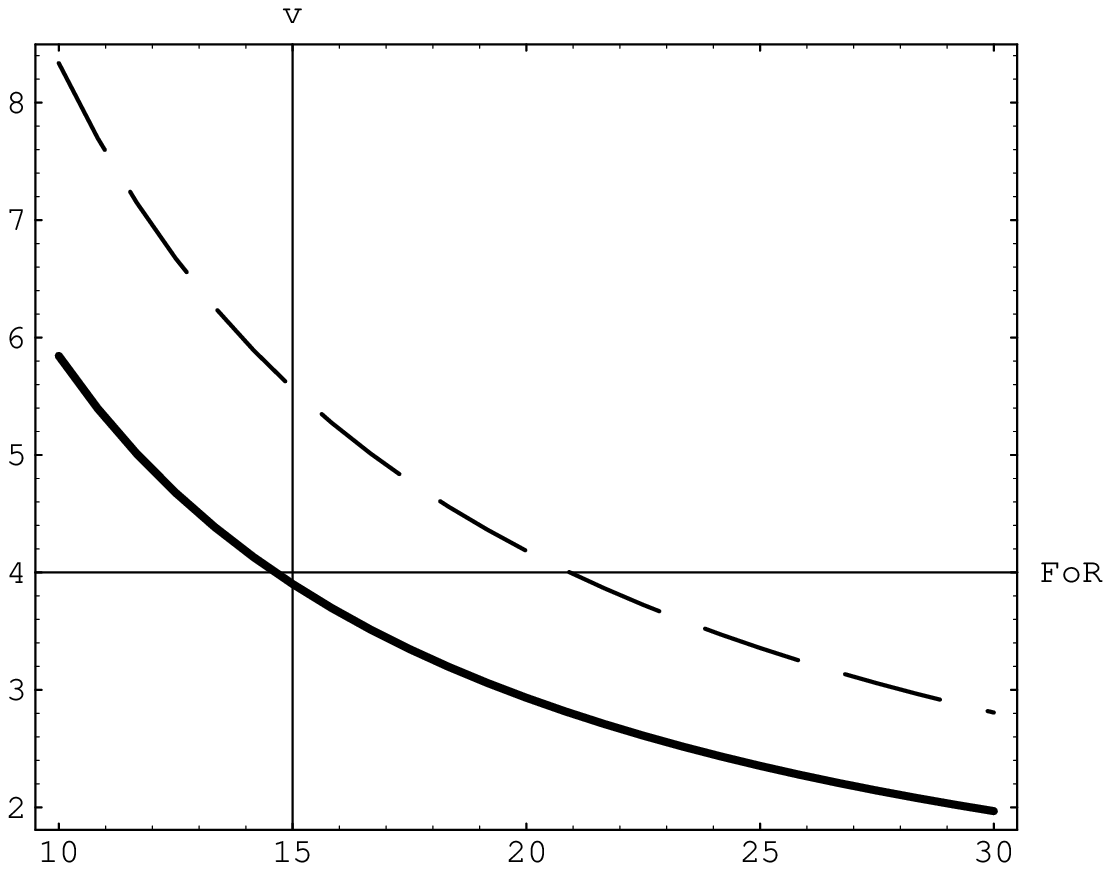,width=100mm,height=90mm}

\rotatebox{0}{\hskip 0pt Size of the field of regard (FoR), ~deg}
     \caption{{The size of the tolerable velocity error as a function of
FoR. Plotted from  Eq.(\ref{eq:simmain3c3}) for the  
astrometric error of  $\sigma_\alpha$ allocated  
for the relativistic stellar aberration in the total error budget.   
The upper dashed curve correspond to  
$\sigma_\alpha=1~\mu$as  and the lower thick curve is for
$\sigma_\alpha=0.701 ~\mu$as.}
      \label{fig:for}}
 \end{center}
\end{figure}
A possible increase of the field of regard (FoR), which is currently being
discussed, will also  results in minimizing $\sigma_v$. In the Figure
\ref{fig:for} we present the dependence  of the maximal tolerable error in
the velocity determination $\sigma_v$ as a function  of FoR.
One can see that relativistic stellar aberration increases roughly as the
tile angle squared.  Thus, increasing the tile size from 15 to 20 degrees
puts a tighter requirement on the knowledge of the spacecraft velocity
which  is already difficult because of the earth trailing orbit.
Therefore a further study of this problem  is granted.

\subsection{Requirements for on-board attitude control}

In this Section we will estimate the influence of a temporal changes
introduced by the rotational motion of the interferometer with respect to
it's center of mass. Thus, the expression (\ref{eq:simmain54}) may be used to
derive another requirement, namely the requirement on the knowledge of the
rotational motion of the spacecraft, or on-board attitude control. A
simple-minded calculation assumes that, for example, the angle
$\psi$ is changing with time as

\begin{equation}
\psi(\tau)=\psi_0+\dot{\psi}_0\cdot \tau +\frac{1}{2}\ddot{\psi}_0\cdot
\tau^2+{\cal O}(\tau^3),
\label{egphi}
\end{equation}
\noindent where $\dot{\psi}_0$ and $\ddot{\psi}_0$ are the constant rate of
this angular change and constant angular acceleration (a
similar analysis could be made for the angle $\epsilon$). 
Assuming that  for any given time the quantity $\psi(\tau)$ has a normal 
Gaussian distribution with errors obeying the equation:  
\begin{equation}
\Delta \psi(\tau)= \Delta\dot{\psi}_0\cdot \tau + {\cal O}(\tau^2).
\label{egdelphi}
\end{equation}
the later expression may be taken  as an input to derive the
on-board attitude  requirements. Let the time interval for sampling  
will be of the order of $\tau=\Delta t ~{\rm  s}$ then from
Eqs.(\ref{eq:simmain4}) and (\ref{egdelphi}) one derives the following 
constraint:
\begin{equation}
\sigma_{\dot{\psi}_0}\sim  
38.529\cdot\frac{\Delta k}{\Delta t}  ~{\rm  mas/s}. 
\label{eqdel274}
\end{equation}
Thus, for the sampling rate of $\tau=0.5 ~{\rm  s}$ (e.q 
$\Delta t =0.5$) and $\Delta k=0.701$ (for absolute
angular acceleration compensation, e.q. $\ddot{\psi}_0=0$), from this last
equation (\ref{eqdel274}) one  obtains the following  requirement on the
accuracy of spacecraft attitude control:
\begin{equation}
\sigma^{[0.7]}_{\dot{\psi}_0}\sim ~ 54.018 ~{\rm  mas/s}. 
\label{eq:del28}
\end{equation}

Concluding, we would  like to mention that the derived requirements
Eqs.(\ref{eq:simmain5}),(\ref{eq:del28}) were based upon the
assumption that the total error budget for the  velocity aberration is
$\sigma_{\alpha_v}=0.701~\mu$as.  Naturally,   minimizing
the error budget for the velocity aberration $\sigma_{\alpha_v}$   will
result in tightening the derived constraints on   the knowledge of
the velocity vector
itself $\vec{v}=v\big(\cos\psi\cos\epsilon,~\sin\psi\cos\epsilon,
~\sin\epsilon\,\big)$ and the corresponding
$\sigma_v,\sigma_\phi,\sigma_\epsilon$.

\section*{Conclusions}

We have demonstrated that the 4 mm/s requirement for the relativistic stellar
aberration is quite well justified for the SIM related studies. However, a
secondary effects, such as  possible correlation between the constant term
and the instantaneous spacecraft velocity  vector may produce a noticeable
impact on the astrometric observations.  Thus, account for this possible
correlation, actually results in  tightening  the  velocity requirement 
(for the worst case of highly correlated data) and almost twice minimizes a
tolerable error in the velocity estimates, notably  $\sigma_v =1.616$ 
mm/s.  Additionally, the solar radiation pressure will greatly influence the
SIM orbital solution and will be the major force acting on the spacecraft. 
As a result the navigation DSN time will not go down as the mission
progresses because of fluctuations in the solar radiation pressure. 

These issues may be addressed by analyzing a number of different options
that will present a grounds for possible   trades between the Mission,
Spacecraft, and Instrument Systems. Some of these  options are 
1). On-board processing of data, as oppose to increasing the use of  
DSN time;  2). The use of a precisely positioned solar shade for maintaining
constant pressure loading;  3). Velocity determination using 
integrated and time averaged accelerometers.
In any case the development of a better model for the instrument in order
to perform analysis of its `elastic'/dynamical properties  and the
systematics hidden in the ``constant'' term will significantly boost the
related studies. 

Another issue that needed to be discussed is the reducing
the velocity accuracy requirement  down to 2 mm/s.  A first look at
this problem brings a conclusion is that this is not an easy problem for SIM
in deep space (for more details, please see \cite{YouEllis98,Gorham98}).
Thus, some of the issues concerning SIM in a SIRTF-like orbit, with a 2mm/s
velocity knowledge requirement are leading to the following conclusions from
the DSN tracking stand-point: 

1. By utilizing the Doppler range-rate method one can achieve more than
adequate precision in the radial direction, of order 0.05 mm/s is
reasonable. But in the plane of the sky, it is more difficult: long
observations (doppler arcs) are necessary and it is unclear that the same
precision can be obtained uniformly in any case -- it will likely be a
function of the ecliptic latitude of SIM at the time of observation.  

2. By using the  Delta-DOR observations (VLBI) with a single baseline and 
X/S-band observations, one may be able to obtain a precision of at best $\sim$ 1
nrad for a fairly short observation (tens of minutes, including QSO calibrator). 
This corresponds to $\sim$ 15 m at 0.1~AU. So, a pair of observations
separated by $\sim$ 2 hr or, so could get one plane-of-sky component to the
required accuracy; ideally a track would be used to improve the SNR and get
somewhat better accuracy perpendicular to the baseline. To get both plane of
sky components a pair of baselines is necessary [a bigger DSN commitment, of
course].

It is unclear whether the radial Doppler range-rate will be available 
with the same antennas at the same time as the VLBI observations. 
Development of a Ka band capability would improve the precision by a
factor of $\sim$ 3. This is why meeting the SIM requirements is not
completely trivial. It might improve if the distance is not as far, and it
would be interesting to see how performance varied with a number
of parameters, including declination. There is another issue which was
raised in this paper. That is a reasonable definition for the data
sampling rate $\Delta t$ necessary to use in order to introduce an on-board 
correction for the   relativistic aberration. Our results show that the answer to
this question may impact the whole Mission operations approach.
This is why we believe that this issue should be a topic for a separate
study and the corresponding results will be reported elsewhere.

Let us mention that the resulting SIM astrometric catalog will be
derived by reducing the collected data to the solar system barycenter. 
In this case the uncertainty in Earth's barycentric velocity may be additional
source of errors. Let us estimate the accuracy of the presently known value for
Earth' barycentric velocity $v_\oplus$. It is known that the Earth velocity
uncertainty is dominated  by the uncertainty in the Earth's orbital plane
orientation,  at the $\sim$1 mas level, giving a velocity uncertainty of
$\sim$0.03 mm/s. However, there may be given another more conservative
estimate to this quantity by using the  accuracy of  determination of  the
Earth-Moon barycenter, which is presently known to accuracy of about $\sim
1~$km.  This number helps to  estimate the  present accuracy of knowledge
the Earth's barycentric velocity as  $\sigma_{v_\oplus}=0.2~$mm/s, thus
offering a fair  possibility for $\sim 1~\mu$as astrometry [that, as we saw,
requires $\sigma_v\sim$ 4mm/s]. So this is not a problem for SIM.

The requirements on the navigational accuracy obtained here, relate to
the interferometer itself and not to the center of mass of the spacecraft. In
the case of changing of the center mass of the spacecraft (which may occur
due to different reasons, such as the rotation during imaging mode, or due to the
use of propellant, etc.), the navigational parameters should be re-calculated to
the required accuracy.  Let us also mention that at the SIM level of accuracy
a more subtle effects on the motion  of the Earth about the sun will start to
play an important role.  These effects are the  Moon's and the largest
planets influence, and, probably, the body-body  interaction between the
Earth and the Moon, the spin-orbital  connection in the Earth dynamics, the
solar rotations and pulsations, influence of solar plasma, etc. All of these
effects will have  to be calculated and properly included into the models for
astrometric grid simulations and corresponding  data analysis.   Our further
studies will be aimed on  improvement of the models used for the presented
analysis by developing a better model for the instrument behavior and by
expanding the parameter space to incorporate  the other  physical phenomena
affecting astrometric observations from within the solar system.  

\acknowledgments
The reported research   has been done at the Jet Propulsion
Laboratory,  California Institute of Technology, which is under  contract to the 
National Aeronautic and Space Administration.


\newpage
\appendix 
\section{\Large\bfseries~ General expression for the errors propagation}
\label{appa}

In this Appendix we will present  the intermediate calculations 
omitted   for brevity in the main text. Parameterization
Eq.(\ref{eq:vect}) 
\begin{eqnarray}
\vec{s}_1&=&\big(\cos\alpha_1\cos\delta_1,~\sin\alpha_1\cos\delta_1,
~\sin\delta_1\big),\nonumber\\[1mm]
\vec{s}_2&=&\big(\cos\alpha_2\cos\delta_2,~\sin\alpha_2\cos\delta_2,
~\sin\delta_2\big),\nonumber\\[1mm]
\vec{v}&=&v\big(\cos\psi\cos\epsilon,~\sin\psi\cos\epsilon,
~\sin\epsilon\,\big),\nonumber\\[1mm]
\vec{b}&=&b\big(\cos\mu\cos\nu,~\sin\mu\cos\nu,~\sin\nu\,\big). 
\label{eq:vect0}
\end{eqnarray}

\noindent allows one to express the scalar products in   
Eq.(\ref{eq:delmain}) in the usual way:
{}
\begin{eqnarray}
(\vec{b}\cdot\vec{s}_1)&=& b\Big(
\cos\nu\cos\delta_1\cos(\alpha_1-\mu)+\sin\nu\sin\delta_1\Big), \nonumber\\ 
(\vec{b}\cdot\vec{s}_2)&=& b\Big(
\cos\nu\cos\delta_2\cos(\alpha_2-\mu)+\sin\nu\sin\delta_2\Big), \nonumber\\ 
(\vec{v} \cdot \vec{s}_1)&=& v\Big(
\cos\epsilon\cos\delta_1\cos(\alpha_1-\psi)+\sin\epsilon\sin\delta_1\Big),
\nonumber\\  
(\vec{v}\cdot\vec{s}_2)&=& v\Big(
\cos\epsilon\cos\delta_2\cos(\alpha_2-\psi)+\sin\epsilon\sin\delta_2\Big),
\label{eq:scalar}
\end{eqnarray}
\noindent As a result, we can present equation (\ref{eq:delmain}) 
in the fully-blown form as follows: {}
\begin{eqnarray}
\delta\ell &=& b\Big[\cos\nu\Big(\cos\delta_1\cos(\alpha_1-\mu)-
\cos\delta_2\cos(\alpha_2-\mu)\Big)+
\sin\nu\Big(\sin\delta_1-\sin\delta_2\Big)\Big]+\delta c_0-\nonumber\\[1.5mm]
&-&\frac{bv}{c}\Big[
\Big(\cos\nu\cos\delta_1\cos(\alpha_1-\mu)+\sin\nu\sin\delta_1\Big)
\Big(\cos\epsilon\cos\delta_1\cos(\alpha_1-\psi)+
\sin\epsilon\sin\delta_1\Big)-\nonumber\\[1.5mm]
&-&\Big(\cos\nu\cos\delta_2\cos(\alpha_2-\mu)+\sin\nu\sin\delta_2\Big)
\Big(\cos\epsilon\cos\delta_2\cos(\alpha_2-\psi)+
\sin\epsilon\sin\delta_2\Big)\Big].
\label{eq:del12} 
\end{eqnarray}

By taking the first derivative from both left and right sides
of this equation (\ref{eq:del}),   we will obtain expression  necessary to
analyze  the  contributions of the different error factors involved in the
problem under consideration to the overall error budget:

\begin{eqnarray}
\Delta\delta\ell &=&
\sum_{i=1,2}\Big(\frac{\partial\delta\ell}{\partial\alpha_i}\Delta\alpha_i 
+\frac{\partial\delta\ell}{\partial\delta_i}\Delta\delta_i\Big)+ 
\frac{\partial\delta\ell}{\partial b}\Delta b+\nonumber\\
&+&\frac{\partial\delta\ell}{\partial \mu}\Delta \mu+
\frac{\partial\delta\ell}{\partial \nu}\Delta \nu+
\frac{\partial\delta\ell}{\partial v}\Delta v+
\frac{\partial\delta\ell}{\partial \epsilon}\Delta\epsilon+
\frac{\partial\delta\ell}{\partial \psi}\Delta\psi+
\Delta\delta c_0. 
\label{eq:del_full}
\end{eqnarray}

\noindent with the partial derivatives with respect to the observing angles
$\delta_1,\delta_2$ and $\alpha_1,\alpha_2$ given as follows:

\begin{eqnarray}
\frac{\partial\delta\ell}{\partial\delta_1}&=&
b~\Big\{-\cos\nu\sin\delta_1\cos(\alpha_1-\mu)
+\sin\nu\cos\delta_1+\nonumber\\  
&+&\frac{v}{c}\Big[\Big(\cos\epsilon
\cos\delta_1\cos(\alpha_1-\psi)+\sin\epsilon\sin\delta_1\Big)
\Big(\cos\nu\sin\delta_1\cos(\alpha_1-\mu)-\sin\nu\cos\delta_1\Big)+\nonumber\\
&&+~\Big(\cos\epsilon\sin\delta_1
\cos(\alpha_1-\psi)-\sin\epsilon\cos\delta_1\Big)
\Big(\cos\nu\cos\delta_1\cos(\alpha_1-\mu)+\sin\nu\sin\delta_1
\Big)\Big]\Big\},\nonumber\\
\frac{\partial\delta\ell}{\partial\delta_2}&=&
b~\Big\{\cos\nu\sin\delta_2\cos(\alpha_2-\mu)-\sin\nu\cos\delta_2-\nonumber\\
&-&\frac{v}{c}\Big[\Big(\cos\epsilon
\cos\delta_2\cos(\alpha_2-\psi)+\sin\epsilon\sin\delta_2\Big)
\Big(\cos\nu\sin\delta_2\cos(\alpha_2-\mu)-\sin\nu\cos\delta_2\Big)+\nonumber\\
&&+~\Big(\cos\epsilon\sin\delta_2
\cos(\alpha_2-\psi)-\sin\epsilon\cos\delta_2\Big)
\Big(\cos\nu\cos\delta_2\cos(\alpha_2-\mu)+\sin\nu\sin\delta_2\Big)
\Big]\Big\},\nonumber\\ 
\frac{\partial\delta\ell}{\partial\alpha_1}&=&
b~\Big\{-\cos\nu\cos\delta_1\sin(\alpha_1-\mu)+\nonumber\\
& &~+~\frac{v}{c}\Big[\Big(\cos\epsilon
\cos\delta_1\cos(\alpha_1-\psi)+\sin\epsilon\sin\delta_1\Big)
\cos\nu\cos\delta_1\sin(\alpha_1-\mu)+
\nonumber\\ 
& &~~~~~+~\Big(\cos\nu\cos\delta_1\cos(\alpha_1-\mu)+
\sin\nu\sin\delta_1\Big)\cos\epsilon
\cos\delta_1\sin(\alpha_1-\psi)\Big]\Big\},\nonumber\\  
\frac{\partial\delta\ell}{\partial\alpha_2}&=&
b~\Big\{\cos\nu\cos\delta_2\sin(\alpha_2-\mu)-\nonumber\\
& &~-~\frac{v}{c}\Big[\Big(\cos\epsilon
\cos\delta_2\cos(\alpha_2-\psi)+\sin\epsilon \sin\delta_2\Big)
\cos\nu\cos\delta_2\sin(\alpha_2-\mu)+\nonumber\\
& & ~~~~~+~\Big(\cos\nu\cos\delta_2\cos(\alpha_2 - \mu)+
\sin\nu\sin\delta_2\Big)\cos\epsilon 
\cos\delta_2\sin(\alpha_2-\psi)\Big]\Big\}, 
\label{eq:partials1} 
\end{eqnarray}

\noindent together with the partial derivatives with respect to components of
the   baseline vector $\vec{b}$:

\begin{eqnarray}
\frac{\partial\delta\ell}{\partial b}&=&
~\cos\nu\Big(\cos\delta_1\cos(\alpha_1-\mu)-
\cos\delta_2\cos(\alpha_2-\mu)\Big)+\sin\nu\Big(\sin\delta_1-\sin\delta_2\Big)+\nonumber\\ 
&+&\frac{v}{c}\Big[\Big(\cos\epsilon\cos\delta_2\cos(\alpha_2-\psi)+
\sin\epsilon\sin\delta_2\Big)
\Big(\cos\nu\cos\delta_2\cos(\alpha_2-\mu)+\sin\nu\sin\delta_2\Big)-\nonumber\\
&&-~\Big(\cos\epsilon\cos\delta_1\cos(\alpha_1-\psi)+
\sin\epsilon\sin\delta_1\Big)
\Big(\cos\nu\cos\delta_1\cos(\alpha_1-\mu)+\sin\nu\sin\delta_1\Big)\Big],
\nonumber\\
\frac{\partial\delta\ell}{\partial \nu}&=&
b~\Big\{\sin\nu\Big(\cos\delta_2\cos(\alpha_2-\mu)-
\cos\delta_1\cos(\alpha_1-\mu)\Big)+\cos\nu\Big(\sin\delta_1-\sin\delta_2\Big)+\nonumber\\ 
&+&\frac{v}{c}\Big[\Big(\cos\epsilon\cos\delta_2\cos(\alpha_2-\psi)+
\sin\epsilon\sin\delta_2\Big)
\Big(-\sin\nu\cos\delta_2\cos(\alpha_2-\mu)+\cos\nu\sin\delta_2\Big)-\nonumber\\
&&-~\Big(\cos\epsilon\cos\delta_1\cos(\alpha_1-\psi)+
\sin\epsilon\sin\delta_1\Big)
\Big(-\sin\nu\cos\delta_1\cos(\alpha_1-\mu)+\cos\nu\sin\delta_1\Big)\Big]\Big\},
\nonumber\\
\frac{\partial\delta\ell}{\partial \mu}&=&
b~\cos\nu \Big\{\cos\delta_1\cos(\alpha_1-\mu)-
\cos\delta_2\cos(\alpha_2-\mu)+\nonumber\\ 
&+&\frac{v}{c}\Big[\Big(\cos\epsilon\cos\delta_2\cos(\alpha_2-\psi)+
\sin\epsilon\sin\delta_2\Big)\cos\delta_2\sin(\alpha_2-\mu)-\nonumber\\
&&-~\Big(\cos\epsilon\cos\delta_1\cos(\alpha_1-\psi)+
\sin\epsilon\sin\delta_1\Big)\cos\delta_1\sin(\alpha_1-\mu)\Big]\Big\},
\label{eq:partials2} 
\end{eqnarray}

\noindent and, finally, with the following partials for the 
spacecraft's velocity:

\begin{eqnarray}
\frac{\partial\delta\ell}{\partial v}&=&
\frac{b}{c}\Big[\Big(\cos\epsilon
\cos\delta_2\cos(\alpha_2-\psi)+\sin\epsilon\sin\delta_2\Big)
\Big(\cos\nu\cos\delta_2\cos(\alpha_2-\mu)+\sin\nu\sin\delta_2\Big)-
\nonumber\\[1.5mm]
& &-~\Big(\cos\epsilon
\cos\delta_1\cos(\alpha_1-\psi)+\sin\epsilon\sin\delta_1\Big)
\Big(\cos\nu\cos\delta_1\cos(\alpha_1-\mu)+\sin\nu\sin\delta_1\Big)
\Big],\nonumber\\ [1.5mm] 
\frac{\partial\delta\ell}{\partial \epsilon}&=&
\frac{bv}{c}\Big[\Big(\sin\epsilon
\cos\delta_1\cos(\alpha_1-\psi)-\cos\epsilon\sin\delta_1\Big)
\Big(\cos\nu\cos\delta_1\cos(\alpha_1-\mu)+\sin\nu\sin\delta_1\Big)-
\nonumber\\ 
& &~-~\Big(\sin\epsilon
\cos\delta_2\cos(\alpha_2-\psi)-\cos\epsilon\sin\delta_2\Big)
\Big(\cos\nu\cos\delta_2\cos(\alpha_2-\mu)+
\sin\nu\sin\delta_2\Big)\Big],\nonumber\\ 
\frac{\partial\delta\ell}{\partial \psi}&=&
 \frac{bv}{c}\cos\epsilon
\Big[\Big(\cos\nu\cos\delta_2\cos(\alpha_2-\mu)+\sin\nu\sin\delta_2\Big)
\cos\delta_2\sin(\alpha_2-\psi)-\nonumber\\
& &~~~~~~~-~\Big(\cos\nu\cos\delta_1\cos(\alpha_1-\mu)+\sin\nu\sin\delta_1\Big)
\cos\delta_1\sin(\alpha_1-\psi)\Big].
\label{eq:partials3} 
\end{eqnarray}

The group of the  expressions Eqs.(\ref{eq:del_full})-(\ref{eq:partials3}) is
quite difficult for analytical description, however it may be significantly
simplified.  First, without loosing generality one can set $\mu=\nu=0$,
which equivalent to choosing the  direction of the baseline vector $\vec{b}$
coinciding with $x$-axis, namely $\vec{b}=b(1,0,0)$.
 As a result of this choice, all vectors   now
will be counted \underline{from the baseline} and the expression
(\ref{eq:del12})   for the relative OPD may  be presented  as follows:
{}
\begin{eqnarray}
\delta\ell &=&b\Big(\cos\delta_1\cos\alpha_1-
\cos\delta_2\cos\alpha_2\Big)+\delta c_0-\nonumber\\[1.5mm]
&-&\frac{bv}{c} 
\Big[\cos\delta_1\cos\alpha_1\, 
\Big(\cos\epsilon\cos\delta_1\cos(\alpha_1-\psi)+
\sin\epsilon\sin\delta_1\Big)-\nonumber\\[1.5mm]
&&~~-\cos\delta_2\cos\alpha_2\, 
\Big(\cos\epsilon\cos\delta_2\cos(\alpha_2-\psi)+
\sin\epsilon\sin\delta_2\Big)\Big].
\label{eq:del} 
\end{eqnarray}
Second,  the largest expected ratio $\frac{v}{c}\approx
\frac{v_\oplus}{c}=9.942\times 10^{-5}$, which  makes  
the  terms $\sim  \frac{v}{c}$ in the expressions (\ref{eq:partials1}) 
and (\ref{eq:partials2}) of the second order of smallness.
 For the purposes of the present
analysis, we may omit these terms and, after some re-arranging, 
 expression  Eq.(\ref{eq:del_full}) may be presented as
\begin{eqnarray}
\frac{\Delta\delta\ell}{b} &=&\Delta\alpha_2\sin\alpha_2\cos\delta_2-
\Delta\alpha_1\sin\alpha_1\cos\delta_1+\nonumber\\[2mm]
&+&\Delta\delta_2\cos\alpha_2\sin\delta_2-
\Delta\delta_1\cos\alpha_1\sin\delta_1+\nonumber\\[2mm] 
&+&\frac{\Delta b}{b}\Big[\cos\alpha_1\cos\delta_1-
\cos\alpha_2\cos\delta_2\Big]+\frac{\Delta\delta c_0}{b}+\nonumber\\[1.5mm] 
&+&
\frac{\Delta v}{c}\Big[\cos\epsilon\,\Big(\cos^2\delta_2\cos\alpha_2 
\cos(\alpha_2-\psi)-\cos^2\delta_1\cos\alpha_1 
\cos(\alpha_1-\psi)\Big)+\nonumber\\[1.5mm]
& &~~+~\sin\epsilon\,\Big(\cos\delta_2\sin\delta_2\cos\alpha_2-
\cos\delta_1\sin\delta_1\cos\alpha_1\Big)\Big]+\nonumber\\[1.5mm]  
&+&\Delta\epsilon~\frac{v}{c}\Big[\sin\epsilon
\,\Big(\cos^2\delta_1\cos\alpha_1 
\cos(\alpha_1-\psi)-\cos^2\delta_2\cos\alpha_2 
\cos(\alpha_2-\psi)\Big)+\nonumber\\[1.5mm]
& &~~~\,+~\cos\epsilon\,\Big(\cos\delta_2\sin\delta_2\cos\alpha_2-
\cos\delta_1\sin\delta_1\cos\alpha_1\Big)\Big]+\nonumber\\ 
&+&\Delta\psi~\frac{v}{c}\cos\epsilon\,
\Big[\cos\alpha_2\cos^2\delta_2\sin(\alpha_2-\psi)-
\cos\alpha_1\cos^2\delta_1\sin(\alpha_1-\psi)\Big].
\label{eqdel15b}
\end{eqnarray}

This last expression may   further be simplified 
by introducing a parameterization, which is natural for the  
problem under consideration. The final result for the 
first derivative of the  differential OPD may be given as follows: 
{}
\begin{eqnarray}
\frac{\Delta\delta\ell}{b} &=&\Delta\alpha \sin\alpha_2\cos\delta_2+
\Delta\alpha_1\Big(\sin\alpha_2\cos\delta_2-
\sin\alpha_1\cos\delta_1\Big)+\nonumber\\[1.5mm]
&+&\Delta\delta \cos\alpha_2\sin\delta_2+
\Delta\delta_1\Big(\cos\alpha_2\sin\delta_2-
\cos\alpha_1\sin\delta_1\Big)+\nonumber\\[1.5mm]
&+&\frac{\Delta b}{b}\Big(\cos\alpha_1\cos\delta_1-
\cos\alpha_2\cos\delta_2\Big)+\frac{\Delta\delta c_0}{b}+\nonumber\\[1.5mm] 
&+&\Big[\frac{\Delta v}{c}\cos(\epsilon-\epsilon_0)
-\Delta\epsilon~\frac{v}{c} \sin(\epsilon-\epsilon_0)\Big]
\sqrt{(a^2+f^2)\cos^2(\psi-\psi_0)+k^2}-\nonumber\\[1.5mm]
&-&\Delta\psi~\frac{v}{c}\cos\epsilon
\sqrt{a^2+f^2}\sin(\psi-\psi_0).
\label{eqdel15ca}
\end{eqnarray}

\noindent where both angles $\psi_0$ and $\epsilon_0$
are depend only on the positions of the two stars involved and 
are given as follows:
\begin{eqnarray}
\sin\psi_0&=&\frac{f}{\sqrt{a^2+f^2}}, ~\qquad~
\cos\psi_0=\frac{a}{\sqrt{a^2+f^2}}, \nonumber\\[2mm]
a&=&\cos^2\delta_2\cos^2\alpha_2 
-\cos^2\delta_1\cos^2\alpha_1, \nonumber\\[2.5mm]
f&=&\cos^2\delta_2\cos\alpha_2\sin\alpha_2-
\cos^2\delta_1\cos\alpha_1\sin\alpha_1.  
\label{eq:psi2} \\[1.5mm]
\sin\epsilon_0&=&\frac{k}{\sqrt{(a^2+f^2)\cos^2(\psi-\psi_0)+k^2}}, 
\nonumber\\[1.5mm]
\cos\epsilon_0&=&\frac{\sqrt{a^2+f^2}\cos(\psi-\psi_0)}
{\sqrt{(a^2+f^2)\cos^2(\psi-\psi_0)+k^2}},\nonumber\\[1.5mm] 
k&=&\cos\delta_2\sin\delta_2\cos\alpha_2 
-\cos\delta_1\sin\delta_1\cos\alpha_1. 
\label{eq:epsilon2}
\end{eqnarray}

\newpage 
\section{\Large\bfseries~ Errors in the fully-parameterized relative OPD}
\label{appb}

Taking into account the fully parameterized form of the fractional 
relative OPD Eqs.(\ref{eq:del_full})-(\ref{eq:partials3}) we may obtain the
form of this quantity for a tile in the SIM nominal observing direction. Thus,
substituting in these expressions the values for the positions of primary  and
secondary stars Eq.(\ref{eq:tile}) we will have:

\begin{eqnarray}
\frac{\Delta\delta\ell}{b}
&=&\Delta\alpha\,\cos\nu\cos\frac{\delta}{2}\cos(\frac{\alpha}{2}-\mu)
+2\Delta(\alpha_1-\mu)\cos\nu\cos\frac{\delta}{2}\sin\frac{\alpha}{2}\sin\mu
-\nonumber\\[2mm]
&-&\Delta\delta\,\Big[\cos\nu\sin\frac{\delta}{2}\sin(\frac{\alpha}{2}-\mu)~+~ 
\sin\nu\cos\frac{\delta}{2}\,\Big]+
2\Delta\delta_1\cos\nu\sin\frac{\delta}{2}\cos\frac{\alpha}{2}\sin\mu-\nonumber\\
&-&2\Delta\nu\Big[\sin\nu\cos\frac{\delta}{2}\sin\frac{\alpha}{2}\cos\mu+
\cos\nu\sin\frac{\delta}{2}\,\Big]+\nonumber\\[2mm]
&+& \frac{2\Delta b}{b}\Big[\cos\nu\cos\frac{\delta}{2}
\sin\frac{\alpha}{2}\cos\mu-\sin\nu\sin\frac{\delta}{2}\Big]
~+~\frac{\Delta\delta c_0}{b}~+\nonumber\\[2mm]
&+&\frac{\Delta v}{c}\Big[\cos\nu\Big(-\cos\epsilon\sin\alpha
\cos^2\frac{\delta}{2}\,\sin(\psi+\mu)+\sin\epsilon\cos\frac{\alpha}{2}
\sin\delta\sin\mu\Big)+\nonumber\\[2mm]
&&~~~~~+\sin\nu\,\cos\epsilon\cos\frac{\alpha}{2}\sin\delta\sin\psi\Big]+
\nonumber\\[2mm]
&+&\Delta\epsilon\,\frac{v}{c}\Big[\cos\nu\Big(\sin\epsilon\sin\alpha
\cos^2\frac{\delta}{2}\,\sin(\psi+\mu)+\cos\epsilon\cos\frac{\alpha}{2}
\sin\delta\sin\mu\Big)-\nonumber\\[2mm]
&&~~~~~-\sin\nu\,\sin\epsilon\cos\frac{\alpha}{2}\sin\delta\sin\psi\Big]+
\nonumber\\[2mm]
&+&\Delta\psi\,\frac{v}{c}\cos\epsilon\,
\Big(\sin\nu\,\cos\frac{\alpha}{2}\sin\delta\cos\psi-
\cos\nu\sin\alpha\cos^2\frac{\delta}{2}\,\cos(\psi+\mu)\Big). 
\label{eq:simmain33}
\end{eqnarray}

One may see that the angles of the baseline orientation ($\mu,\nu$) 
are significantly influencing the narrow-angle astrometric observations.
However, as previously we  will choose both angles as $\mu=\nu=0$,
which is  equivalent to choosing the  direction of the baseline vector
$\vec{b}$ to coincide with $x$-axis, namely $\vec{b}=b(1,0,0)$. 
 As a result of this choice, all vectors   now will be counted from the
baseline (see Figure \ref{fig:differential_astrometry}). Resulted
expression for the relative OPD  may  be presented in a simpler  form,
namely:

\begin{eqnarray}
\frac{\Delta\delta\ell}{b} 
&=&\Delta\alpha\,\cos\frac{\alpha}{2}\cos\frac{\delta}{2}~-~
\Delta\delta\,\sin\frac{\alpha}{2}\sin\frac{\delta}{2}~+~ 
\frac{2\Delta b}{b}\,\sin\frac{\alpha}{2}\cos\frac{\delta}{2}
~+~\frac{\Delta\delta c_0}{b}~-\nonumber\\[2mm]
&-&\Big[\Big(\frac{\Delta v}{c}\,\cos\epsilon
~-~\Delta\epsilon~\frac{v}{c}\,\sin\epsilon\Big)
\sin\psi~+~\Delta\psi~\frac{v}{c}\,\cos\epsilon\cos\psi\Big]
\sin\alpha\cos^2\frac{\delta}{2}. 
\label{eq:simmain22}
\end{eqnarray}

\noindent Note that obtained expression does not depend on the errors in 
position of the primary star $\Delta\alpha_1,\Delta\delta_1$. The obtained
result  may now be used to analyze the  propagation of errors in the future 
astrometric  observations with SIM. 

\newpage 
\section{\Large\bfseries~ Components  of the covariance matrix}
\label{appc}  

The first term in the expression (\ref{eq:simmain22}) 
$\Delta\delta\ell$ may equivalently be presented as
$\Delta\delta\ell=\Delta(n\lambda_0)=
\Delta n\,\lambda_0+n\,\Delta\lambda_0$, where $\lambda_0$  is the operating
frequency and $n$ is an integer number. This term  vanishes because 
both $\lambda_0$ and $n$ ~are assumed to be known to a sufficiently 
high accuracy, ~e.q. $\Delta n=0, ~\Delta\lambda_0=0$. This is true due to
the fact that the SIM white light fringe tracker will make sure that
variations in these two quantities will be exactly zero. 
Then the remaining differentials $\Delta v, ~\Delta\epsilon, ~\Delta\psi,
~\Delta\alpha,  ~\Delta\delta, ~\Delta b $ and
$\Delta\delta c_0$ will have to satisfy the  equation:   
{} 
\begin{eqnarray} 
\Delta\alpha\,\cos\frac{\alpha}{2}\cos\frac{\delta}{2}&-&
\Delta\delta\,\sin\frac{\alpha}{2}\sin\frac{\delta}{2}~=~
-\frac{\Delta\delta c_0}{b}-
\frac{2\Delta b}{b}\,\sin\frac{\alpha}{2}\cos\frac{\delta}{2}
+\nonumber\\[2mm]
+\Big[\Big(\frac{\Delta v}{c}\,\cos\epsilon
&-&\Delta\epsilon~\frac{v}{c}\,\sin\epsilon\Big)
\sin\psi ~+~ \Delta\psi~\frac{v}{c}\,\cos\epsilon\cos\psi\Big]
\sin\alpha\cos^2\frac{\delta}{2}.  
\label{eq:simm1}
\end{eqnarray}
\noindent In order to simplify   further analysis we have separated the 
terms in the expression above in a such a way, so that the left side of this
equation represents the error in the measurement of the absolute angular
separation  between the two stars,  while the left side  shows the main
contributing factors to this quantity.

In order to study  the  error propagation  in the astrometric observations,
we need to take the square of the expression (\ref{eq:simm1}):
{}
\begin{align} 
(\Delta\alpha)^2&\,\cos^2\frac{\alpha}{2}\cos^2\frac{\delta}{2}+ 
(\Delta\delta)^2\,\sin^2\frac{\alpha}{2}\sin^2\frac{\delta}{2}-
\frac{(\Delta\alpha\Delta\delta)}{2}\,\sin\alpha\sin\delta ~=
\nonumber\\[2mm] 
&=~\frac{(\Delta\delta c_0)^2}{b^2}~+~
\frac{4(\Delta b)^2}{b^2}\,\sin^2\frac{\alpha}{2}\,\cos^2\frac{\delta}{2}
~+~4\frac{(\Delta\delta c_0\Delta b)}{b^2}
\,\sin\frac{\alpha}{2}\cos\frac{\delta}{2}+\nonumber\\[2mm]
&+~\Big[\Big(\frac{\Delta v}{c}\,\cos\epsilon
~-~\Delta\epsilon~\frac{v}{c}\,\sin\epsilon\Big)
\sin\psi ~+~ \Delta\psi~\frac{v}{c}\,\cos\epsilon\cos\psi\Big]^2
\sin^2\alpha\cos^4\frac{\delta}{2}-\nonumber\\[2mm]
&-~\frac{2\Delta\delta c_0}{b}\Big[\Big(
\frac{\Delta v}{c}\,\cos\epsilon ~-~
\Delta\epsilon~\frac{v}{c}\,\sin\epsilon\Big)
\sin\psi ~+~ \Delta\psi~\frac{v}{c}\,\cos\epsilon\cos\psi\Big]
\sin\alpha\cos^2\frac{\delta}{2}-\nonumber\\[2mm]
&-~\frac{4\Delta b}{b}\Big[\Big(\frac{\Delta v}{c}\,\cos\epsilon
~-~\Delta\epsilon~\frac{v}{c}\,\sin\epsilon\Big)
\sin\psi ~+~ \Delta\psi~\frac{v}{c}\,\cos\epsilon\cos\psi\Big]
\sin\alpha\sin\frac{\alpha}{2}\cos^3\frac{\delta}{2}.  
\label{eq:simm2}
\end{align}

One may expect that in any given tile the errors in $\Delta\alpha$ and
$\Delta\delta$ are normally distributed and 
uncorrelated.\footnote{Note that this is not true for a general case of
studying the reference grid stability. Thus one finds that the correlation 
in {\tt RA} and {\tt DEC} becomes a source for the zonal errors in the 
analysis of the grid accuracy.}   The errors due to the  orbital dynamics
$\Delta v, ~\Delta\epsilon,  ~\Delta\psi$  at the chosen approximation may
also be treated as being not correlated with  instrumental errors 
$\Delta b, ~\Delta c_0$. As a result, the components of the covariance 
matrix may be  given as follows:
 
\begin{eqnarray}
\overline{(\Delta\alpha)^2}&=&\sigma^2_\alpha, ~~\qquad~~ 
\overline{(\Delta \delta)^2}~=~\sigma^2_\delta, ~~\qquad~~
\overline{\Delta\alpha\Delta\delta}~=~0, \nonumber\\[2mm] 
\overline{(\Delta v)^2} &=& \sigma^2_v, ~~\qquad~~\,
\overline{(\Delta\epsilon)^2}~=~\sigma^2_\epsilon, ~~\qquad~~\,
\overline{(\Delta\psi)^2}~=~\sigma^2_\psi, \nonumber\\[2mm]
\overline{\Delta v\Delta\epsilon}&=&0, ~~~\qquad~~
\overline{\Delta v\Delta\psi}~=~0, ~~~~\qquad~~
\overline{\Delta\epsilon\Delta \psi}~=~0,\nonumber\\[2mm] 
\overline{(\Delta \delta c_0)^2}&=&\sigma^2_{\delta_{c_0}}, \qquad
\overline{\Delta \delta_{c_0}\Delta v}~=~0, ~~~~~\qquad~~ 
\overline{(\Delta b)^2}~=~\sigma^2_b, \nonumber\\[2mm]
\overline{\Delta b\Delta v}&=&0, ~~~\,\qquad~~ 
\overline{\Delta b\Delta\epsilon}~~=~0, ~~~~\qquad~~
\overline{\Delta b\Delta\psi}~=~0.
\label{eq:simm3a}
\end{eqnarray}
{}
\begin{equation}
\overline{\Delta\delta c_0\Delta b}=
\sigma_{\delta_{c_0}}\sigma_b\,\rho(c_0,b),
\qquad
\overline{\Delta \delta c_0\Delta \epsilon}=
\sigma_{\delta_{c_0}}\sigma_\epsilon\,\rho(c_0,\epsilon),
\qquad
\overline{\Delta \delta c_0\Delta\psi}=
\sigma_{\delta_{c_0}}\sigma_\psi\,\rho(c_0,\psi). 
\label{eq:simm3b}
\end{equation}

These expressions are helpful to present the result of averaging the 
equation (\ref{eq:simm2}) in the  following form:
{}
\begin{eqnarray}
&&\sigma^2_\alpha\,\cos^2\frac{\alpha}{2}\cos^2\frac{\delta}{2}+
\sigma^2_\delta\,\sin^2\frac{\alpha}{2}\sin^2\frac{\delta}{2}=\nonumber\\
&&~~~=~\frac{\sigma^2_{\delta_{c_0}}}{b^2}+
\frac{4\,\sigma^2_b}{b^2}\,\sin^2\frac{\alpha}{2}\,\cos^2\frac{\delta}{2}
+\frac{4\,\sigma_{\delta_{c_0}}\sigma_b}{b^2} \rho(c_0,b)
\,\sin\frac{\alpha}{2}\cos\frac{\delta}{2}~+~\nonumber\\[2mm]
&&~~~+~\Big[\Big(\frac{\sigma^2_v}{c^2}\,\cos^2\epsilon
+\sigma^2_\epsilon~\frac{v^2}{c^2}\,\sin^2\epsilon\Big)\sin^2\psi
+\sigma^2_\psi~\frac{v^2}{c^2}\,\cos^2\epsilon\,\cos^2\psi\Big]
\sin^2\alpha\cos^4\frac{\delta}{2}+\nonumber\\
&&~~~+~\frac{2\,\sigma_{\delta_{c_0}}}{b}
\Big[\sigma_\epsilon \,\rho(c_0,\epsilon)\,\sin\epsilon
\sin\psi ~-~ \sigma_\psi\,\rho(c_0,\psi)\,\cos\epsilon\cos\psi\Big]
\,\frac{v}{c}\,\sin\alpha\cos^2\frac{\delta}{2}.
\label{eq:simm4a}
\end{eqnarray}

Expression (\ref{eq:simm4a}) suggests that for the observations in the
direction perpendicular to the baseline (e.q. when $\delta$ are non-zero), 
there will be   a  large contribution to the error budget coming from
$\sigma_\delta$.   
However, for  observations parallel to the baseline (e.q $\delta=0$) 
the influence of $\sigma_\delta$ is vanishes and one obtains the most 
stringent requirement on the velocity aberration. Then, by taking $\delta=0$
one may present Eq.(\ref{eq:simm4a}) in the following form:
 
\begin{eqnarray}
&&\sigma^2_\alpha\,\cos^2\frac{\alpha}{2}=
~\frac{\sigma^2_{\delta_{c_0}}}{b^2}+
\frac{4\,\sigma^2_b}{b^2}\,\sin^2\frac{\alpha}{2}
+\frac{4\,\sigma_{\delta_{c_0}}\sigma_b}{b^2} \rho(c_0,b)
\,\sin\frac{\alpha}{2}~+~\nonumber\\[2mm]
&&~~~+~\Big[\Big(\frac{\sigma^2_v}{c^2}\,\cos^2\epsilon
+\sigma^2_\epsilon~\frac{v^2}{c^2}\,\sin^2\epsilon\Big)\sin^2\psi
+\sigma^2_\psi~\frac{v^2}{c^2}\,\cos^2\epsilon\,\cos^2\psi\Big]
\sin^2\alpha+\nonumber\\
&&~~~+~\frac{2\,\sigma_{\delta_{c_0}}}{b}
\Big[\sigma_\epsilon \,\rho(c_0,\epsilon)\,\sin\epsilon
\sin\psi ~-~ \sigma_\psi\,\rho(c_0,\psi)\,\cos\epsilon\cos\psi\Big]
\,\frac{v}{c}\,\sin\alpha.
\label{eq:simm4b}
\end{eqnarray}
\noindent Remembering further that angles ($\alpha,\delta$)  
vary in the range given by $\alpha^2+\delta^2\leq(\frac{\pi}{12})^2$ and
$\cos\frac{\alpha}{2}$ never vanishes in this interval, we can divide the
both sides of the equation (\ref{eq:simm4b}) on $\cos^2\frac{\alpha}{2}$:
\begin{eqnarray}
&&\sigma^2_\alpha=
~\frac{\sigma^2_{\delta_{c_0}}}{b^2\,\cos^2\frac{\alpha}{2}}+
\frac{4\,\sigma^2_b}{b^2}\,\tan^2\frac{\alpha}{2}
+\frac{4\,\sigma_{\delta_{c_0}}\sigma_b}{b^2} \rho(c_0,b)
\,\frac{\sin\frac{\alpha}{2}}{\cos^2\frac{\alpha}{2}}~+~\nonumber\\[2mm]
&&~~~+~4\Big[\Big(\frac{\sigma^2_v}{c^2}\,\cos^2\epsilon
+\sigma^2_\epsilon~\frac{v^2}{c^2}\,\sin^2\epsilon\Big)\sin^2\psi
+\sigma^2_\psi~\frac{v^2}{c^2}\,\cos^2\epsilon\,\cos^2\psi\Big]
\,\sin^2\frac{\alpha}{2}+\nonumber\\
&&~~~+~\frac{4\,\sigma_{\delta_{c_0}}}{b}
\Big[\sigma_\epsilon \,\rho(c_0,\epsilon)\,\sin\epsilon
\sin\psi ~-~ \sigma_\psi\,\rho(c_0,\psi)\,\cos\epsilon\cos\psi\Big]
\,\frac{v}{c}\,\tan\frac{\alpha}{2}.
\label{eq:simm5}
\end{eqnarray}

This equation represents an ellipsoid of with half-axis defined by 
the direction of motion of the spacecraft. Thus, for the spacecraft motion in
the direction given by $\psi=\frac{\pi}{2}$ and $\epsilon=0$ we can have the
first  expression for astrometric errors:
 
\begin{eqnarray}
&&\sigma^2_\alpha=
~\frac{\sigma^2_{\delta_{c_0}}}{b^2\,\cos^2\frac{\alpha}{2}}+
\frac{4\,\sigma^2_b}{b^2}\,\tan^2\frac{\alpha}{2}
+\frac{4\,\sigma_{\delta_{c_0}}\sigma_b}{b^2} \rho(c_0,b)
\,\frac{\sin\frac{\alpha}{2}}{\cos^2\frac{\alpha}{2}}~+~
\frac{4\,\sigma^2_v}{c^2} \,\sin^2\frac{\alpha}{2}.
\label{eq:simm6}
\end{eqnarray}
 
\noindent For the case of motion in the direction  $\psi=\frac{\pi}{2}$ and
$\epsilon=\pm\frac{\pi}{2}$ we have the second relation: 
 
\begin{eqnarray}
&&\sigma^2_\alpha=
~\frac{\sigma^2_{\delta_{c_0}}}{b^2\,\cos^2\frac{\alpha}{2}}~+~
\frac{4\,\sigma^2_b}{b^2}\,\tan^2\frac{\alpha}{2}
~+~\frac{4\,\sigma_{\delta_{c_0}}\sigma_b}{b^2} \rho(c_0,b)
\,\frac{\sin\frac{\alpha}{2}}{\cos^2\frac{\alpha}{2}}~+~\nonumber\\[2mm]
&&~~~+~4\,\sigma^2_\epsilon~\frac{v^2}{c^2}\,\sin^2\frac{\alpha}{2}~\pm~
\frac{4\,\sigma_{\delta_{c_0}}}{b}\sigma_\epsilon
\,\frac{v}{c} \,\rho(c_0,\epsilon)\,\tan\frac{\alpha}{2}.
\label{eq:simm6b}
\end{eqnarray}
\noindent And, finally, the last expression may be obtained when considering the
motion in the direction $\psi=0$ and $\epsilon=0, \,\pi$. This last constraint
reads: {}
\begin{eqnarray}
&&\sigma^2_\alpha=
~\frac{\sigma^2_{\delta_{c_0}}}{b^2\,\cos^2\frac{\alpha}{2}}+
\frac{4\,\sigma^2_b}{b^2}\,\tan^2\frac{\alpha}{2}
~+~\frac{4\,\sigma_{\delta_{c_0}}\sigma_b}{b^2} \rho(c_0,b)
\,\frac{\sin\frac{\alpha}{2}}{\cos^2\frac{\alpha}{2}}~+~\nonumber\\[2mm]
&&~~~+~4\,\sigma^2_\psi~\frac{v^2}{c^2}
\,\sin^2\frac{\alpha}{2}~\mp~\frac{4\,\sigma_{\delta_{c_0}}}{b}
\sigma_\psi \,\frac{v}{c} \,\rho(c_0,\psi) \,\tan\frac{\alpha}{2}.
\label{eq:simm6c}
\end{eqnarray}

\noindent For the simplicity, we can relate the errors in the sky angles of
velocity $\sigma_\epsilon$ and $\sigma_\psi$ with that of the velocity
magnitude $\sigma_v$. To obtain the most stringent constraints on the sky angle
we will use to following  expressions:
\begin{eqnarray}
\frac{\,\sigma^2_v}{c^2}\,\sin^2\frac{\alpha}{2}&=&
\,\sigma^2_\epsilon~\frac{v^2}{c^2}\,\sin^2\frac{\alpha}{2}~+~
\sigma_\epsilon\,\frac{v}{c}\, \frac{\,\sigma_{\delta_{c_0}}}{b}
\,\rho(c_0,\epsilon)\,\tan\frac{\alpha}{2}~= \nonumber\\
&=&\,\sigma^2_\psi~\frac{v^2}{c^2}
\,\sin^2\frac{\alpha}{2}~+~\sigma_\psi\,\frac{v}{c} 
\,\frac{\,\sigma_{\delta_{c_0}}}{b}\,\rho(c_0,\psi)\,\tan\frac{\alpha}{2}.
\label{eq:simm7}
\end{eqnarray}

As a result, in our further analysis we will concentrate  on the
equation Eq.(\ref{eq:simm6}) together with the following two solutions for 
$\sigma_\epsilon$ and $\sigma_\psi$ (which were obtained directly from
Eqs.(\ref{eq:simm7})):
\begin{eqnarray}
\sigma_\epsilon&=&\sqrt{\frac{\,\sigma^2_v}{v^2}+
\Big[\frac{\,\sigma_{\delta_{c_0}}}{b}
\frac{\,\rho(c_0,\epsilon)}{\,\sin\alpha}\,\frac{c}{v}\Big]^2}
-\frac{\,\sigma_{\delta_{c_0}}}{b}
\frac{\,\rho(c_0,\epsilon)}{\,\sin\alpha}\,\frac{c}{v}~\ge~0, \nonumber\\[2mm]
\sigma_\psi&=&\sqrt{\frac{\,\sigma^2_v}{v^2}+
\Big[\frac{\,\sigma_{\delta_{c_0}}}{b}
\frac{\,\rho(c_0,\psi)}{\,\sin\alpha}\,\frac{c}{v}\Big]^2}
-\frac{\,\sigma_{\delta_{c_0}}}{b}
\frac{\,\rho(c_0,\psi)}{\,\sin\alpha}\,\frac{c}{v}~\ge~0.
\label{eq:angles}
\end{eqnarray}

Just for the estimation purposes we can assume that 
$\rho(c_0,\psi)=\rho(c_0,\epsilon)=\rho_0$. Then, by using the  
expressions  (\ref{eq:angles}) together with the estimates
Eqs.(\ref{eq:simmain3c3}) one could obtain improved requirements on the 
accuracy of knowledge of the two sky-angles of the spacecraft velocity:

\begin{equation}
\sigma_\epsilon= \sigma_\psi
=\sigma_\alpha\,\frac{c}{v}\,\frac{1}{2\sin\frac{\alpha}{2}}
\Big(\sqrt{1+\rho^2_0}-\rho_0\Big)\ge0.
\end{equation}
\noindent Note, that for the worst case of highly correlated quantities
($\psi,\epsilon$) and $c_0$, e.q. $\rho_0\sim1$ the result
Eq.(\ref{eq:simmain5}) will have to be further reduced by a factor
$\sqrt{2}-1=0.4142$, thus tightening the  requirements for the accuracy of
knowledge of the  two velocity sky-angles  as $\sigma_\epsilon =
\sigma_\psi= 20.186$ mas. 
This example suggests that a possible correlation between the velocity
components and the constant term may be reduced if the  DSN
will deliver the data for all three components with a twice higher
accuracy then planned.

\end{document}